\documentclass[english]{article}
\usepackage[T1]{fontenc}
\usepackage[latin9]{inputenc}
\usepackage{geometry}
\usepackage{float}
\usepackage{amsmath}
\usepackage{amssymb}
\usepackage{graphicx}
\usepackage{stmaryrd}

\makeatletter

\providecommand{\tabularnewline}{\\}

\PassOptionsToPackage{normalem}{ulem} 
\usepackage{babel}
\usepackage{cite}
\usepackage{color}
\usepackage{dsfont}
\usepackage{empheq}
\usepackage{framed}
\usepackage{mathrsfs}
\usepackage{needspace}
\usepackage{setspace}
\usepackage{tabularx}
\usepackage{tikz}
\usepackage{ulem}
\usepackage{url}

\let\@fnsymbol\@arabic

\newcommand{\mm}{\mu_{\mathrm{macro}}}
\newcommand{\lm}{\lambda_{\mathrm{macro}}}
\newcommand{\mh}{\mu_{\mathrm{micro}}}
\newcommand{\lh}{\lambda_{\mathrm{micro}}}
\newcommand{\me}{\mu_{e}}
\newcommand{\mc}{\mu_{c}}
\newcommand{\lle}{\lambda_{e}}

\newcommand{\mLc}{\me L_{c}^{2}}

\newcommand{\nablau}{\,\nabla u\,}
\newcommand{\p}{{P}}
\newcommand{\nablap}{\nabla \p}
\newcommand{\Curl}{\,\mathrm{Curl}}
\newcommand{\dev}{\, \mathrm{dev}}
\newcommand{\Div}{\mathrm{Div}}
\newcommand{\tr}{\, \mathrm{tr}}
\newcommand{\sym}{\, \mathrm{sym}\,}

\renewcommand{\skew}{\, \mathrm{skew}\,}

\newcommand{\x}{\cdot}

\newcommand{\sig}{\widetilde{\sigma}}
\newcommand{\Sig}{\sigma}
\newcommand{\n}{ n}

\renewcommand{\skew}{\, \mathrm{skew}}

\newcommand{\devsym}{\dev\sym}

\newcommand{\id}{\,\mathds{1}}

\definecolor{Green}{rgb}{0,0.52,0}

\newcommand*{\widefbox}[1]{\fbox{\hspace{2em}#1\hspace{2em}}}

\title{\vspace{-1.0cm}Relaxed micromorphic modeling of the interface between a homogeneous solid and a band-gap metamaterial: new perspectives towards meta-structural design}

\author{
	Angela Madeo\footnote{Angela Madeo, corresponding author, angela.madeo@insa-lyon.fr, Univ Lyon, INSA-Lyon, SMS-ID, 20 avenue	Albert Einstein, F-69621, Villeurbanne cedex and IUF, Institut universitaire de France, 1 rue Descartes, 75231, Paris Cedex 05, France}\, and  
Gabriele Barbagallo\footnote{Gabriele Barbagallo, gabriele.barbagallo@insa-lyon.fr, Univ Lyon, INSA-Lyon, SMS-ID, LaMCoS, CNRS, 20 avenue Albert Einstein, F-69621, Villeurbanne cedex, France} \, and 
 Manuel Collet\footnote{Manuel Collet, manuel.collet@ec-lyon.fr, Ecole Centrale de Lyon, LTDS UMR-CNRS 5513, 36 avenue Guy de Collongue, 69134, Ecully, France} \, and \\
 Marco Valerio d'Agostino\footnote{Marco Valerio d'Agostino, marco-valerio.dagostino@insa-lyon.fr, Univ Lyon, INSA-Lyon, SMS-ID, 20 avenue Albert Einstein, F-69621, Villeurbanne cedex, France} \, and
	Marco Miniaci\footnote{Marco Miniaci, marco.miniaci@gmail.com, Universit\'e du Havre, Laboratoire Ondes et Milieux Complexes, UMR CNRS 6294, 75 Rue Bellot, 76600, Le Havre, France} \, and 
Patrizio Neff\,\footnote{Patrizio Neff, patrizio.neff@uni-due.de, Head of Chair for Nonlinear Analysis and Modelling, Fakultät für Mathematik, Universität Duisburg-Essen,  Mathematik-Carrée, Thea-Leymann-Straße 9, 45127 Essen}  
}

\makeatother

\usepackage{babel}
\begin{document}
\maketitle 
\addtocounter{footnote}{4} \vspace{0.5cm}
 
\begin{abstract}
In the present paper, the material parameters of the isotropic relaxed micromorphic model derived for a specific metamaterial in a previous contribution  are used to model its transmission properties. Specifically, the reflection and transmission coefficients at an interface between a homogeneous solid and the chosen metamaterial are analyzed by using both the relaxed micromorphic model and a direct FEM implementation of the detailed microstructure. The obtained results show an excellent agreement between the transmission spectra derived via our enriched continuum model and those issued by the direct FEM simulation. Such excellent agreement validates the indirect measure of the material parameters and opens the way towards an efficient meta-structural design.
\end{abstract}
\vspace{0.5cm}

\hspace{-0.55cm}\textbf{Keywords}: reflection, transmission, gradient micro-inertia, free micro-inertia,
complete band-gaps, non-local effects, relaxed micromorphic model,
generalized continuum models, meta-structures

\vspace{0.4cm}

\hspace{-0.55cm}\textbf{AMS 2010 subject classification}: 74A10 (stress),
74A30 (nonsimple materials), 74A60 (micromechanical theories), 74E15
(crystalline structure), 74M25 (micromechanics), 74Q15 (effective
constitutive equations)

\vspace{0.5cm}


\tableofcontents{}\vspace{1.2cm}
 

\section{Introduction}

Metamaterials are engineered composite structures exhibiting unconventional properties due to a specific spatial organization of their single constituents. For instance, it is possible to create materials that have the ability to inhibit elastic wave propagation in specific frequency ranges usually called frequency band gaps \cite{deymier2013acoustic}. To obtain this result, it is possible to exploit two main kinds of phenomena that are strongly related to the presence of an underlying microstructure. The first type is based on a local resonance (Mie resonance) of a micro-deformation mode that absorbs the energy of propagating waves. These micro-oscillations confine the motion in the micro-structure impeding any macroscopic propagation. Alternatively, the second kind of phenomena is based on some micro-diffusion phenomena (Bragg scattering): the propagating wave has wavelengths comparable to the dimension of the microstructure leading to reflection and transmission phenomena at the micro-level that globally inhibit macroscopic wave propagation. The frequency values of the band gap and all other related characteristics depend on the metamaterial microstructural properties, such as topology organization and single constituents stiffness.

\medskip{}

The common approach to the mechanical description of metamaterials is that of introducing models that take into account all the details of the underlying microstructures  (see e.g. \cite{liu2000locally,pham2013transient,sridhar2016homogenization,spadoni2009phononic,aurialt2012long,boutin2003homogenisation,boutin2010generalized,boutin2011generalized}). Such detailed approaches, while having the advantage of faithfully reproducing the microstructural topologies, may not be adapted for envisaging the description of the chosen metamaterials at the scale of the engineering piece or at the scale of the structure. Indeed, to proceed towards the design of metastructures (structures which are made of metamaterials as building blocks) some averaged models, which allow for a simplified description of the mechanical behavior of metamaterials, are needed.

In other words, we propose to use our simplified enriched continuum model to design morphologically complex metastructures without the need of coding all the details of metamaterials' microstructure. This idea is analogous to the most commonly used approach in classical structural mechanics: it is not needed to account for the type and position of all the atoms inside the considered materials to describe their average behavior and to conceive a simplified approach for the design of complex structures. Such simplified approach usually follows the lines of classical Cauchy continuum mechanics.

Cauchy continuum theory is a powerful tool for structural design involving homogeneous materials. Indeed, in the isotropic setting, it is sufficient to know the value of two constants, namely the Young modulus and the Poisson ratio in order to completely master the mechanical behavior of a specific material. Such constants are true material ``constants'' in the sense that the choice of their specific values is sufficient to completely describe the deformation of the corresponding material subjected to given macroscopic loading conditions. As a matter of fact, such simplified continuum approach is so effective that it allows today the conception of morphologically complex civil and aeronautical structures in optimized finite element environments.

The idea that we propose in this paper follows the same line. Indeed, we introduce an enriched continuum model that is able to account for the averaged behavior of metamaterials via the introduction of few extra coefficients in addition to the classical Young modulus and Poisson ratio. The model that we propose to use to accomplish this goal is the relaxed micromorphic model \cite{madeo2014band,madeo2016reflection}. The pertinence of such model for the mechanical characterization of band-gap metamaterials was proven in previous works \cite{madeo2016first,madeo2016modeling}. Such model indeed proved its effectiveness for the description of dispersion curves in non-local band-gap metamaterials, compared against any other enriched continuum model \cite{madeo2016complete}.

In \cite{madeo2016modeling}, we provided the conclusive proof of the fitting of the relaxed micromorphic model on an actual metamaterial on the basis of a rather simplified procedure based on the comparison of the dispersion curves obtained with a Bloch-wave analysis and with the relaxed micromorphic model. The parameters that we derived in \cite{madeo2016modeling} for a specific metamaterial are material parameters, which means that they are completely descriptive of the mechanical behavior of the considered metamaterial at the macroscopic scale. This means that if we now consider a more complex structure constituted by such metamaterial, the parameters that have been previously identified must be able to describe the behavior of the metastructure in a more complex setting. To test this idea, we consider in the present paper a rather simplified meta-structure consisting of 40 unit cells of the metamaterial characterized in \cite{madeo2016modeling} and we study its reflective properties when such structure is connected to a homogeneous plate.

We underline the fact that such restriction on the complexity of the structure is not dictated by the relaxed micromorphic model, but by the computational time of the detailed FEM simulation that we used to validate our values of the metamaterial parameters. Indeed, for such a rather simplified structure, the detailed FEM model takes almost 20 minutes to compute the solution, while the solution found via the relaxed micromorphic model needs a computational time of only 1-2 minutes. The FEM computational time ulteriorly increases when the mesh is refined around particular frequency values (e.g. around frequencies at which transmission peaks occur due to internal resonances of the microstructure).

The relaxed micromorphic model used in \cite{madeo2016modeling} features a kinetic energy that includes both free and gradient micro-inertia. Both those micro-inertia terms have been shown to be essential for a complete description of the dispersion properties of  metamaterials \cite{madeo2016modeling,madeo2017role,madeo2016review}.

We want to underline again that the material parameters of the relaxed micromorphic model derived in \cite{madeo2016complete} are true material constants in the sense that they are completely representative of the mechanical behavior of that specific metamaterial. Such constants are thus fixed once the metamaterial is fixed and they describe the dynamical behavior of the chosen metamaterial for a wide range of frequencies and for wavelengths which go down to the size of the unit cell. Unlike other homogenized constants that may be determined via the currently used homogenization techniques, the material parameters of the relaxed micromorphic model do not depend  on the frequency. 

Finally, we mention the fact that, being purely macroscopic, the relaxed micromorphic model could be also used for the description of the behavior of other completely different types of band-gap metamaterials as those manufactured using piezoelectric components \cite{fan2016energy,yi2016flexural}. In this last case, the physical meaning of the micro-distortion tensor $P$ should be re-interpreted to be linked to the piezoelectrical fields.

\medskip{}

The present paper is organized according to the following structure:
\begin{itemize}
	\item In section \ref{relaxed},  the considered relaxed micromorphic model is presented providing the strain and kinetic energy, the associated equations of motion and boundary conditions, and an expression for the energy flux derived from the principle of conservation of energy,
	\item In section \ref{Plane}, the hypothesis of plane wave is introduced, which gives rise to simplified equations of motion. The dispersion relations obtained by means of our relaxed micromorphic model are obtained following what done in \cite{madeo2017role}. 
	\item In section \ref{Reflection}, the reflection and transmission coefficients which describe the reflective properties of an interface between a Cauchy material and a relaxed micromorphic continuum are introduced generalizing the results presented in \cite{madeo2016reflection}.	 
	\item In section \ref{FEM}, the metamaterial presented in \cite{madeo2016modeling} is considered. The transmission spectra are determined with a FEM model and with the relaxed micromorphic model using the parameters determined in \cite{madeo2016modeling}. The obtained results are compared showing a very good correspondence.
\end{itemize}

\section{The relaxed micromorphic model with weighted free and gradient micro-inertia \label{relaxed}}

In this section, we recall the main properties of the relaxed micromorphic continuum model presented in \cite{madeo2015wave,neff2014unifying} with the addition of the gradient micro-inertia term proposed in \cite{madeo2017role,madeo2016review} and the decomposition of the free micro-inertia \cite{dagostino2017panorama}. Following \cite{madeo2016reflection}, we derive the expression of the energy flux needed for the computation of the reflection and transmission coefficients.
As shown in \cite{madeo2016modeling}, the relaxed micromorphic model accounting for both weighted free and gradient micro-inertia is able to catch the dispersion patterns of real band-gap metamaterials with few constitutive coefficients that, in contrast to what usually happens when using homogenization techniques, do not depend on frequency. The proposed model has proven its effectiveness for a wide range of frequencies and wavelengths which can also become comparable to the size of the unit cell. With respect to the classical Cauchy continua, the kinematics of micromorphic media is enriched with supplementary kinematical fields related to the presence of a deformable microstructure. This enriched kinematics influences the overall mechanical behavior of the considered continuum allowing to model microscopic-related effects. In a dynamical analysis, the motions of the extended kinematics reflect physical phenomena such as local resonances in which the energy is trapped in local vibrating modes.

\subsection{Strain and kinetic energy densities}

The relaxed micromorphic model endows Mindlin-Eringen's representation with the second order \textbf{dislocation density tensor} $\alpha=-\Curl\p$ instead of the full gradient $\nablap$.\footnote{The dislocation tensor is defined as $\alpha_{ij}=-\left(\Curl\p\right)_{ij}=-\p_{ih,k}\epsilon_{jkh}$, where $\epsilon$ is the Levi-Civita tensor and Einstein notation of sum over repeated indexes is used.} In the isotropic case, the elastic strain energy density reads
\begin{align}
W= & \underbrace{\me\,\lVert\sym\left(\nablau-\p\right)\rVert^{2}+\frac{\lle}{2}\left(\mathrm{tr}\left(\nablau-\p\right)\right)^{2}}_{\mathrm{{\textstyle isotropic\ elastic-energy}}}+\hspace{-0.1cm}\underbrace{\mc\,\lVert\skew\left(\nablau-\p\right)\rVert^{2}}_{\mathrm{{\textstyle rotational\ elastic\ coupling}}}\hspace{-0.1cm}\label{Energy}\\
 & \quad+\underbrace{\mh\,\lVert\sym\p\rVert^{2}+\frac{\lh}{2}\,\left(\mathrm{tr}\p\right)^{2}}_{\mathrm{{\textstyle micro-self-energy}}}+\hspace{-0.2cm}\underbrace{\frac{\mLc}{2}\,\lVert\Curl\p\rVert^{2}}_{\mathrm{{\textstyle isotropic\ curvature}}}\,,\nonumber 
\end{align}
where the parameters and the elastic stress are fromally analogous to the standard Mindlin-Eringen micromorphic model. The relaxed micromorphic model counts 6 constitutive parameters in the isotropic case ($\me$, $\lle$, $\mh$, $\lh$, $\mc$, $L_{c}$). The characteristic length $L_{c}$ is intrinsically related to non-local effects due to the fact that it weights a suitable combination of first order space derivatives in the strain energy density \eqref{Energy}. For a general presentation of the features of the relaxed micromorphic model in the anisotropic setting, we refer to \cite{barbagallo2017transparent}. The model is well-posed in the statical and dynamical case including when $\mc=0$, see \cite{neff2015relaxed,ghiba2014relaxed}.

In the relaxed model the complexity of the general micromorphic model has been decisively reduced featuring basically only symmetric gradient micro-like variables and the $\Curl$ of the micro-distortion $\p$. However, the relaxed model is still general enough to include the full micro-stretch as well as the full Cosserat micro-polar model and the micro-voids model, see \cite{neff2014unifying}. Furthermore, well-posedness results for the statical and dynamical cases have been provided in \cite{neff2014unifying} making decisive use of recently established new coercive inequalities, generalizing Korn's inequality to incompatible tensor fields  \cite{neff2015poincare,neff2012maxwell,neff2011canonical,bauer2014new,bauer2016dev}.

In a dynamic regime, the relaxed micromorphic model has been proven to be the only continuum model that is simultaneously able to describe band-gaps and non-localities in mechanical metamaterials.

As for the kinetic energy, we consider in this paper that it takes the following form (see \cite{madeo2017role,madeo2016modeling}): 
\begin{align}
J=\hspace{-0.1cm}\underbrace{\frac{1}{2}\rho\left\Vert u_{,t}\right\Vert ^{2}}_{\text{Cauchy inertia}}+\hspace{-0.1cm}\underbrace{\frac{1}{2}\eta_{1}\left\Vert \dev\sym\p_{,t}\right\Vert ^{2}+\frac{1}{2}\eta_{2}\left\Vert \skew\p_{,t}\right\Vert ^{2}+\frac{1}{6}\eta_{3}\tr\left(\p_{,t}\right)^{2}}_{\text{weighted free micro-inertia}},\label{Kinetic}\\\nonumber  +\hspace{0.1cm}\underbrace{\frac{1}{2}\overline{\eta}_{1}\left\Vert \dev\sym\nablau_{,t}\right\Vert ^{2}+\frac{1}{2}\overline{\eta}_{2}\left\Vert \skew\nablau_{,t}\right\Vert ^{2}+\frac{1}{6}\overline{\eta}_{3}\tr\left(\nablau_{,t}\right)^{2}}_{\text{gradient micro-inertia}},
\end{align}
where $\rho$ is the value of the average macroscopic mass density
of the considered metamaterial, $\eta_{i},i=\{1,2,3\}$ are the free micro-inertia
densities associated to the different terms of the Cartan-Lie decomposition of $\p$ and the $\overline{\eta}_{i},i=\{1,2,3\}$ are the gradient
micro-inertia densities associated to the different terms of the Cartan-Lie
decomposition of $\nablau$.

In the first part of this paper, we will present the results obtained using the numerical values of the elastic coefficients used as in Table \ref{ParametersValues} if not differently specified. The values assumed by the free micro-inertiae and the gradient micro-inertiae will be given in the captions. The scope of the first numerical simulations being purely illustrative, the chosen values have to be interpreted as tentative values selected with the aim of describing some characteristic effects of the parameters of the relaxed micromorphic  model either on the dispersion curves or on the transmission coefficient.

\begin{table}[H]
	\begin{centering}
		\begin{tabular}{|c|c|c|}
			\hline 
			Parameter  & Value  & Unit\tabularnewline
			\hline 
			\hline 
			$\me$  & $200$  & MPa\tabularnewline
			\hline 
			$\lle=2\me$  & 400  & MPa\tabularnewline
			\hline 
			$\mc=5\me$  & 1000  & MPa\tabularnewline
			\hline 
			$\mh$  & 100  & MPa\tabularnewline
			\hline 
			$\lh$  & $100$  & MPa\tabularnewline
			\hline 
			$L_{c}\ $  & $1$  & mm\tabularnewline
			\hline 
			$\rho$  & $2000$  & kg/m$^{3}$\tabularnewline
			\hline 
		\end{tabular}\quad{}\quad{}\quad{}\quad{}%
		\begin{tabular}{|c|c|c|}
			\hline 
			Parameter  & Value  & Unit\tabularnewline
			\hline 
			\hline 
			$\lm$  & $82.5$  & MPa\tabularnewline
			\hline 
			$\mm$  & $66.7$  & MPa\tabularnewline
			\hline 
			$E_{\mathrm{macro}}$  & $170$  & MPa\tabularnewline
			\hline 
			$\nu_{\mathrm{macro}}$  & $0.28$  & $-$\tabularnewline
			\hline 
		\end{tabular}
		\par\end{centering}
	
	\caption{\label{ParametersValues}Values of the parameters used in the numerical
		simulations (left) and corresponding values of the Lam\'e parameters
		and of the Young modulus and Poisson ratio (right). For the formulas
		needed to calculate the homogenized macroscopic parameters starting
		from the microscopic ones, see \cite{barbagallo2017transparent}.}
\end{table}

\subsection{Equation of motion and boundary conditions}
The equations of motion of the relaxed micromorphic model associated to the strain energy density \eqref{Energy} and to the kinetic energy density \eqref{Kinetic} can be obtained by a classical least action principle and take the form (see \cite{madeo2016reflection,madeo2015wave,neff2017real,neff2015relaxed,madeo2014band,madeo2017role})
\begin{align}
\rho\,u_{,tt}-\hspace{-0,8cm}\underbrace{\Div[\,\mathcal{I}\,]}_{\text{new augmented term}}\hspace{-0,8cm} & =\Div\left[\,\widetilde{\sigma}\,\right], &  \widetilde{\mathcal{I}}& =\widetilde{\sigma}-s-\Curl\,m,&\forall x\in\Omega,\label{eq:Dyn}
\end{align}
where 
\begin{align}
\mathcal{I} & =\overline{\eta}_{1}\,\dev\sym\nablau_{,tt}+\overline{\eta}_{2}\,\skew\nablau_{,tt}+\frac{\overline{\eta}_{3}}{3}\,\tr\left(\nablau_{,tt}\right), \nonumber\\\label{quantities}
\widetilde{\mathcal{I}}&=\eta_1\,\devsym \p_{,tt}+\eta_2\,\skew \p_{,tt}+\frac{\eta_3}{3}\,\tr \left(\p_{,tt}\right),\\\nonumber
\widetilde{\sigma} & =2\,\me\,\sym\left(\nablau-\p\right)+\lle\,\tr\left(\nablau-\p\right)\id+2\,\mc\,\skew\left(\nablau-\p\right),\\
s & =2\,\mh\,\sym\p+\lh\,\tr\left(\!\p\right)\id,\nonumber \\
m & =\mLc\,\Curl\p.\nonumber 
\end{align}
The system of equations in $\p$ can be split into their $\dev \sym$, $\skew$ and $\tr$ parts obtaining (see \cite{dagostino2017panorama}):
\begin{align}
\rho\,u_{,tt}-\Div[\,\mathcal{I}\,] & =\Div\left[\,2\,\me\,\sym\left(\nablau-\p\right)+\lle\,\tr\left(\nablau-\p\right)\id+2\,\mc\,\skew\left(\nablau-\p\right)\,\right],\nonumber \\
\eta_{1}\,\dev\sym\,\p_{,tt} & =2\,\mu_{e}\,\dev\sym\,\left(\nabla u-P\right)-2\,\mh\dev\sym\,P-\mLc\,\dev\sym\left(\Curl\,\Curl\,P\right), \label{SystemSplit} \\\nonumber
\eta_{2}\skew\,\p_{,tt} & =2\,\mu_{c}\skew\,\left(\nabla u-P\right)-\mLc\,\skew\left(\Curl\,\Curl\,P\right), \\\nonumber
\frac{1}{3}\eta_{3}\tr\left(\p_{,tt}\right) & =\left(\frac{2}{3}\mu_{e}+\lambda_{e}\right)\tr\left(\nabla u-P\right)-\left(\frac{2}{3}\mh+\lh\right)\tr \left(\p\right)-\frac{1}{3}\mLc\tr\left(\Curl\,\Curl\,P\right).
\end{align}
The fact of adding a gradient micro-inertia in the kinetic energy \eqref{Kinetic} modifies the strong-form PDEs of the relaxed micromorphic model with the addition of the new term $\mathcal{I}$, see \cite{madeo2017role}. Of course, the associated natural and kinematical boundary conditions are also modified with respect to the ones presented in \cite{madeo2016reflection,madeo2016first} giving:
\begin{align}
t&:=\big(\,\sig+\hspace{-1.1cm}\underbrace{\mathcal{I}}_{\text{new augmented term}}\hspace{-1.1cm}\big)\x \n =t^{\mathrm{ext}}\hspace{-1cm} &\mathrm{or} \qquad\quad&\,u=u_{0},& \quad &\forall x\in\partial\Omega,\label{Boundary}\\
\tau\cdot\nu_i&:= -m\x\epsilon\x\n= \tau^{\mathrm{ext}} \hspace{-1cm} &\mathrm{or}  \qquad\quad&\p\x \nu_{i}=p_{i}, \qquad \qquad i=2,3, & \quad &\forall x\in\partial\Omega,\nonumber
\end{align}
where $\epsilon$ is the Levi-Civita third order tensor, $n$ is the normal to the boundary, $\nu_1$ and $\nu_2$ are 2 orthogonal vectors tangent to the boundary, and $u_0,p_2,p_3,t^{\mathrm{ext}},\tau^{\mathrm{ext}}$ are assigned quantities.
	
On the other hand, if one considers an internal surface, one can find a set of jump duality conditions that can be imposed at surfaces of discontinuity of the material properties in relaxed media which reads 
\begin{align}
\left\llbracket \:\left\langle t\,,\delta u\right\rangle \:\right\rrbracket &=0,&\hspace{-2cm}\left\llbracket \:\left\langle \tau\,,\,\delta\p\right\rangle \:\right\rrbracket& =0. \label{Jump}
\end{align}
There are several possible ways of connecting two micromorphic media in such a way that conditions \eqref{Jump} are verified. Indeed,thinking to classical structural mechanics, it is immediate to understand that the same structural elements can be interconnected using different constraints (for example, in beam 	theory, one can deal with clamps, pivots, rollers, etc.). The reasoning in the case of generalized continua gets more complex. We do not show here how different micromorphic elements can be connected together in such a way that the duality jump conditions \eqref{Jump} are verified, but we adress the reader to \cite{madeo2016reflection}. Here, the only difference with respect to \cite{madeo2016reflection} is the new definition of the force $t$ given in \eqref{Boundary}, where the additional term $\mathcal{I}$ appears. The type of connection that will be used in this paper for simulating a certain coupling between a Cauchy and a micromorphic medium is the "internal clamp with free microstructure" that will be recalled in subsection \ref{Interface}. Such type of constraint has been seen to be useful for the description of real interfaces between classical materials and metamaterials \cite{madeo2016first}.

\subsection{Conservation of total energy and energy flux}
It is known that if  conservative mechanical systems are considered, like in the present paper, then the conservation of total energy must be verified in the form
\begin{align}
E_{,t} +\mathrm{div}\, H = 0,
\end{align}

where $E = J +W$ is the total energy of the system and $H$ is the energy flux vector. It is clear that the explicit expressions for the total energy and for the energy flux are different depending on whether one considers a classical Cauchy model or a relaxed micromorphic model. If the expression of the total energy $E$ is straightforward for the two mentioned cases (it suffices to look at the given expressions of $J$ and $W$), the explicit expression of the energy flux $H$ can be more complicated to be obtained. In our case, $H$ takes the form:
\begin{align}
H=-\left[\widetilde{\sigma}+\mathcal{I}\right]^{T} \cdot u_{,t}-(m^T \cdot \p_{,t}) : \epsilon, \label{Flux}
\end{align}
where the stress tensor $\widetilde{\sigma}$ and the hyper-stress tensor $m$ have been defined in \eqref{quantities} in terms of the basic kinematical field. The proof is analogous to the one presented in \cite{madeo2016reflection} and it is shown in the Appendix.

We also recall that for Cauchy continua:
\begin{align}
H=-\sigma \cdot u_{,t}, \label{FluxCauchy}
\end{align}
where $\sigma$ is the Cauchy stress
\begin{align}
\sigma & =2\,\me\,\sym\left(\nablau\right)+\lle\,\tr\left(\nablau\right)\id.
\end{align}

\section{Plane wave propagation \label{Plane}}

Sufficiently far from a source, dynamic wave solutions may be treated
as planar waves. Therefore, we now want to study harmonic solutions
traveling in an infinite domain for the differential system \eqref{SystemSplit}.
We suppose that the space dependence of all introduced kinematical
fields are limited to the scalar component $X$ which is also the
direction of wave propagation.

With this simplifying hypothesis, the system \eqref{SystemSplit} can be rewritten as:
\begin{itemize}
\item a set of three equations only involving longitudinal quantities: 
\begin{align}
\rho\,\ddot{u}_{1}-\frac{2\,\overline{\eta}_{1}+\overline{\eta}_{3}}{3}\,\ddot{u}_{1,11} & =\left(2\,\me+\lle\right)u_{1,11}-2\me\,P_{,1}^{D}-(2\me+3\lle)\,P_{,1}^{S}\,,\vspace{0.4cm}\nonumber \\
\eta_1\,\ddot{P}^{D} & =\frac{4}{3}\,\me\,u_{1,1}+\frac{1}{3}\,\mLc\,P_{,11}^{D}-\frac{2}{3}\,\mLc P_{,11}^{S}-2\left(\me+\mh\right)\,P^{D}\,,\vspace{0.4cm}\label{Long}\\
\eta_3\,\ddot{P}^{S} & =\frac{2\,\me+3\,\lle}{3}\,u_{1,1}-\frac{1}{3}\,\mLc P_{,11}^{D}+\frac{2}{3}\,\mLc P_{,11}^{S}\nonumber \\
& \quad-\left(2\,\me+3\,\lle+2\,\mh+3\,\lh\right)\,P^{S}\,,\nonumber 
\end{align}

\item two sets of three equations only involving transverse quantities in
the $\xi$-th direction, with $\xi=2,3$: 
\begin{align}
\rho\,\ddot{u}_{\xi}-\frac{\,\overline{\eta}_{1}+\overline{\eta}_{2}}{2}\,\ddot{u}_{\xi,11} & =\left(\me+\mc\right)u_{\xi,11}-2\,\me\,P_{\left(1\xi\right),1}+2\,\mc\,P_{\left[1\xi\right],1},\vspace{0.4cm}\nonumber \\
\eta_1\,\ddot{P}_{\left(1\xi\right)} & =\me\,u_{\xi,1}+\frac{1}{2}\,\mLc\,P_{(1\xi)}{}_{,11}+\frac{1}{2}\,\mLc\,P_{\left[1\xi\right],11}\label{Trans}\\
 & \quad-2\left(\me+\mh\right)\,P_{(1\xi)},\vspace{0.4cm}\nonumber \\
\eta_2\,\ddot{P}_{\left[1\xi\right]} & =-\mc\,u_{\xi,1}+\frac{1}{2}\,\mLc\,P_{(1\xi),11}+\frac{1}{2}\,\mLc P_{\left[1\xi\right]}{}_{,11}-2\,\mc\,P_{\left[1\xi\right]},\nonumber 
\end{align}

\end{itemize}

\begin{itemize}
\item one equation only involving the variable $P_{\left(23\right)}$: 
\begin{align}
\eta_1\,\ddot{P}_{\left(23\right)}=-2\left(\me+\mh\right)P_{\left(23\right)}+\mLc P_{\left(23\right),11},\label{Shear}
\end{align}

\item one equation only involving the variable $P_{\left[23\right]}$ :
\begin{align}
\eta_2\,\ddot{P}_{\left[23\right]}=-2\,\mc\,P_{\left[23\right]}+\mLc P_{\left[23\right],11},\label{Rotations23}
\end{align}

\item one equation only involving the variable $P^{V}$: 
\begin{align}
\eta_1\,\ddot{P}^{V}=-2\left(\me+\mh\right)P^{V}+\mLc P_{,11}^{V},\label{VolumeVariation}
\end{align}
\end{itemize}
where we defined (see also \cite{dagostino2017panorama})
\begin{align}
\p^{S} & :=\frac{1}{3}\tr\left(\p\right), & \p_{\left[ij\right]} & :=\left(\skew\p\right)_{ij}=\frac{1}{2}\left(\p_{ij}-\p_{ji}\right),\label{Decom}\\
\p^{D} & :=\p_{11}-\p^{S}, & \p_{\left(ij\right)} & :=\left(\sym\p\right)_{ij}=\frac{1}{2}\left(\p_{ij}+\p_{ji}\right),\nonumber \\
P^{V} & :=P_{22}-P_{33}.\nonumber 
\end{align}

\subsection{Compact representation of the equations of motion}
To express in compact form the obtained equations we define:
\begin{align}
 v_{1}=\left(u_{1},P^{D},P^{S}\right),\qquad v_{\tau}=\left(u_{\tau},P_{(1\tau)},P_{[1\tau]}\right),\quad\tau=2,3,\qquad v_{4}=\left(P_{(23)},P_{[23]},P^{V}\right). \label{Variables}
\end{align}
Considering this new definition, we now look for a wave form solution of the type: 
\begin{equation}
\underbrace{ v_{1}(X,t)=\boldsymbol{\beta}\,e^{i(kX-\omega\,t)}}_{\text{longitudinal}},\qquad\underbrace{ v_{\tau}(X,t)=\boldsymbol{\gamma}\,^{\tau}e^{i(kX-\omega\,t)}}_{\text{transversal}},\quad\tau=2,3,\qquad\underbrace{ v_{4}(X,t)=\boldsymbol{\gamma}\,^{4}e^{i(kX-\omega\,t)}}_{\text{uncoupled}},\label{WaveForm2}
\end{equation}
where $\boldsymbol{\beta}=(\beta_{1},\beta_{2},\beta_{3})^{T}\in\mathbb{C}^{3}$,
$\boldsymbol{\gamma}^{\tau}=(\gamma_{1}^{\tau},\gamma_{2}^{\tau},\gamma_{3}^{\tau})^{T}\in\mathbb{C}^{3}$
and $\boldsymbol{\gamma}^{4}=(\gamma_{1}^{4},\gamma_{2}^{4},\gamma_{3}^{4})^{T}\in\mathbb{C}^{3}$
are the unknown amplitudes of the considered waves\footnote{Here, we understand that having found the (in general, complex) solutions
of \eqref{WaveForm2} only the real or imaginary parts separately
constitute actual wave solutions which can be observed in reality.}, $k$ is the wavenumber and $\omega$ is the wave-frequency.

Replacing the wave form solution \eqref{WaveForm2} in Eqs. \eqref{Long},
\eqref{Trans}, \eqref{Shear}, \eqref{Rotations23} and \eqref{VolumeVariation},
it is possible to express the system as: 
\begin{equation}
 A_{1}\cdot\boldsymbol{\beta}=0,\qquad\qquad A_{\tau}\cdot\boldsymbol{\gamma}^{\tau}=0,\quad\tau=2,3,\qquad\qquad A_{4}\cdot\boldsymbol{\gamma}^{4}=0,\label{AlgSys}
\end{equation}
where 

\begin{align}
A_{1}(\omega,k)\, & =\left(\begin{array}{ccc}
-\omega^{2}\left(1+k^{2}\,\frac{2\,\overline{\eta}_{1}+\overline{\eta}_{3}}{3\,\rho}\right)+c_{p}^{2}\,k^{2} & \,i\:k\,2\,\me/\rho\  & i\:k\:\left(2\,\me+3\,\lle\right)/\rho\\
\\
-i\:k\,\frac{4}{3}\,\me/\eta_{1} & -\omega^{2}+\frac{1}{3}k^{2}c_{m1}^{2}+\omega_{s}^{2} & -\frac{2}{3}\,k^{2}c_{m1}^{2}\\
\\
-\frac{1}{3}\,i\,k\:\left(2\,\me+3\,\lle\right)/\eta_{3} & -\frac{1}{3}\,k^{2}\,c_{m3}^{2} & -\omega^{2}+\frac{2}{3}\,k^{2}\,c_{m3}^{2}+\omega_{p}^{2}
\end{array}\right),\nonumber \\
\nonumber \\
A_{2}(\omega,k)=A_{3}(\omega,k)\, & =\left(\begin{array}{ccc}
-\omega^{2}\left(1+k^{2}\,\frac{\overline{\eta}_{1}+\overline{\eta}_{2}}{2\,\rho}\right)+k^{2}c_{s}^{2}\  & \,i\,k\,2\,\me/\rho\  & -i\,\frac{\eta_{2}}{\rho}\,\omega_{r}^{2}k\,,\\
\\
-\,i\,k\,\me/\eta_{1}, & -\omega^{2}+\frac{1}{2}\,c_{m1}^{2}\,k^{2}+\omega_{s}^{2} & \frac{1}{2}\,c_{m1}^{2}\,k^{2}\\
\\
\frac{1}{2}\,i\,\omega_{r}^{2}\,k & \frac{1}{2}\,c_{m2}^{2}\,k^{2} & -\omega^{2}+\frac{1\,}{2}c_{m2}^{2}\,k^{2}+\omega_{r}^{2}
\end{array}\right),\nonumber \\ \label{Matrices}
\\
A_{4}(\omega,k)\, & =\left(\begin{array}{ccc}
-\omega^{2}+c_{m1}^{2}\,k^{2}+\omega_{s}^{2} & 0 & 0\\
\\
0 & -\omega^{2}+c_{m2}^{2}\,k^{2}+\omega_{r}^{2} & 0\\
\\
0 & 0 & -\omega^{2}+c_{m1}^{2}\,k^{2}+\omega_{s}^{2}
\end{array}\right).\nonumber 
\end{align}

In the definition of the matrices $A_{i}$, $i=\{1,2,3,4\}$ the following
characteristic quantities have also been introduced:
\begin{empheq}[box=\widefbox]{align*}
\omega_{s} & =\sqrt{\frac{2\,(\mu_{e}+\mh)}{\eta_{1}}},\qquad\omega_{r}=\sqrt{\frac{2\,\mu_{c}}{\eta_{2}}},\qquad\omega_{p}=\sqrt{\frac{(3\,\lambda_{e}+2\,\mu_{e})+(3\,\lh+2\,\mh)}{\eta_{3}}},\vspace{1.2mm}\nonumber \\
\label{eq:characteristic_quantities}\\
c_{m1} & =\sqrt{\frac{\mu_{e}\,L_{c}^{2}}{\eta_{1}}},\qquad c_{m2}=\sqrt{\frac{\mu_{e}\,L_{c}^{2}}{\eta_{2}}},\qquad c_{m3}=\sqrt{\frac{\mu_{e}\,L_{c}^{2}}{\eta_{3}}}\qquad\ \ c_{p}=\sqrt{\frac{\lambda_{e}+2\mu_{e}}{\rho}},\qquad c_{s}=\sqrt{\frac{\mu_{e}+\mu_{c}}{\rho}}.\nonumber 
\end{empheq}

In order to have non-trivial solutions of the algebraic systems (\ref{AlgSys}),
one must impose that 
\begin{equation}
\underbrace{\mathrm{det}\, A_{1}(\omega,k)=0,}_{\text{longitudinal}}\qquad\qquad\underbrace{\mathrm{det}\, A_{2}(\omega,k)=\mathrm{det}\, A_{3}(\omega,k)=0,}_{\text{transverse}}\qquad\qquad\underbrace{\mathrm{det}\, A_{4}(\omega,k)=0,}_{\text{uncoupled}}\label{Dispersion}
\end{equation}
The solutions $\omega=\omega(k)$ of these algebraic equations are called the dispersion curves of the relaxed micromorphic model for longitudinal, transverse and uncoupled waves, respectively. Since the explicit expressions of the dispersion curves are rather complex we do not report them here, while showing their behavior in Figure \ref{EtaFullLong}.

\subsection{Dispersion relations\label{sec:Dispersion}}

\begin{figure}[H]
\begin{centering}
\includegraphics[width=6cm]{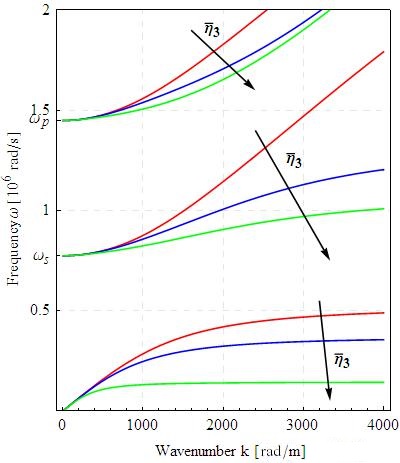} \hspace{2cm}
\includegraphics[width=6cm]{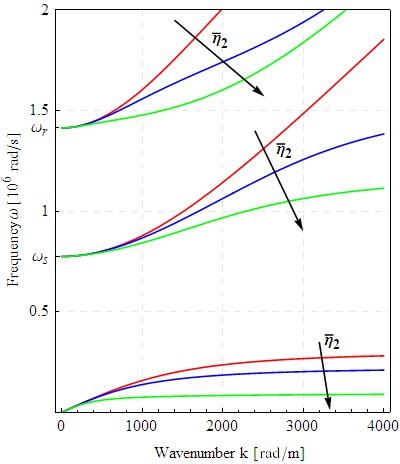} 
\par\end{centering}

\caption{\label{EtaFullLong}Dispersion relations $\omega=\omega(k)$ of the
\textbf{relaxed micromorphic model} for the longitudinal waves with
free micro-inertia $\eta_1=\eta_2=\eta_3=10^{-3}$ and gradient micro-inertia $\overline{\eta}_{3}=(3\times10^{-4},3\times10^{-3},3\times10^{-2})$ kg/m
(left) and transverse waves with micro-inertia $\eta=10^{-3}$ and
gradient micro-inertia $\overline{\eta}_{2}=(2\times10^{-4},2\times10^{-3},2\times10^{-2})$ kg/m
(right). We also set $\overline{\eta}_{1}=0$ since this parameter simultaneously acts on longitudinal and transverse waves which is not a desired effect.}
\end{figure}

Figure \ref{EtaFullLong} shows the dispersion curves obtained for
non-null free micro-inertia $\eta_i\neq0$ with the addition of gradient
micro-inertia $\overline{\eta}_i\neq0$. Surprisingly, as shown in \cite{madeo2017role} the combined effect of the traditional free micro-inertia $\eta_i$ with the gradient micro-inertiae $\overline{\eta}_i$ can lead to the onset of a second longitudinal and transverse band gap. Moreover, it is possible to notice that the addition of gradient micro-inertiae $\overline{\eta}_{1}$, $\overline{\eta}_{2}$ and $\overline{\eta}_{3}$ has no effect on the cut-off frequencies, which only depend on the free micro-inertia $\eta_i$ (and of course
on the constitutive parameters).  

The uncoupled waves in the relaxed micromorphic model with generalized micro-inertia behave as in the classical relaxed micromorphic model (see \cite{madeo2014band,madeo2015wave}) as it is possible
to see analyzing the matrix $A_4$ in Equation \eqref{Matrices}.

\begin{figure}[H]
	\begin{centering}
		\includegraphics[height=6.2cm]{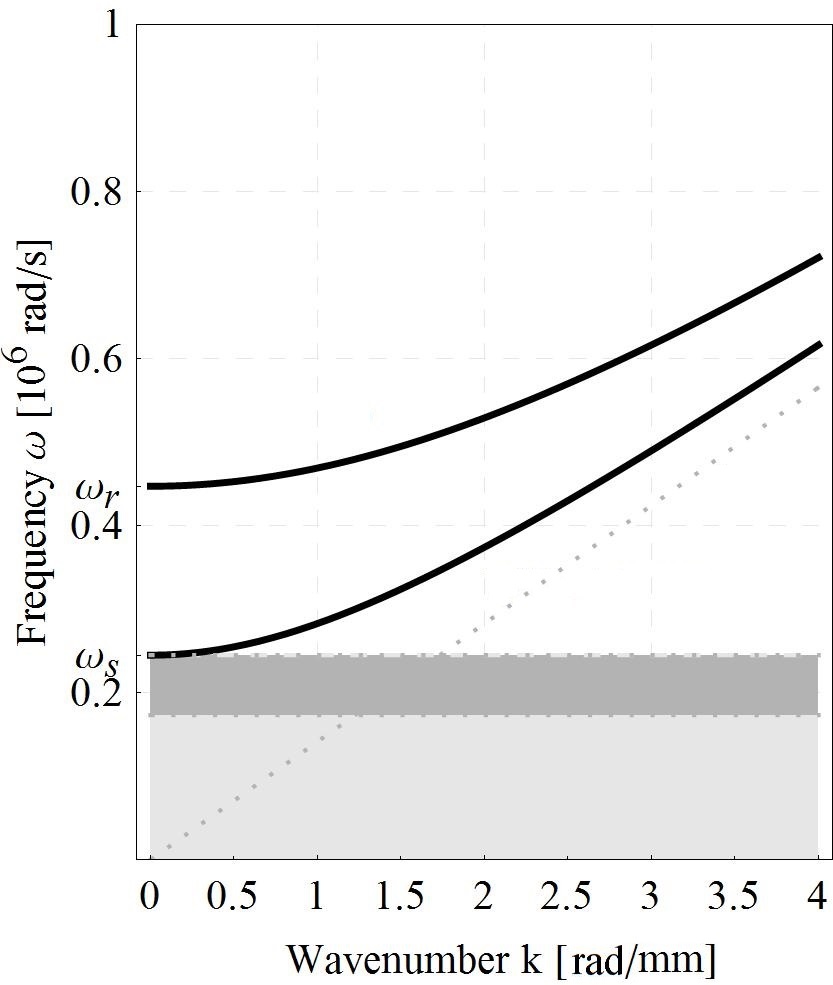}
		\par\end{centering}
	
	\caption{Dispersion relations $\omega=\omega(k)$ for the uncoupled waves of the \textbf{relaxed
			micromorphic model} with free micro-inertia $\eta_1=\eta_2=\eta_3=10^{-2}$ kg/m. The behavior of these curves is not affected at all by any possible form of gradient micro-inertia. \label{Relaxed} }
\end{figure}

The resulting dispersion curves are the same to the ones obtained
with the classical relaxed micromorphic model, see Figure \ref{Relaxed}. For a study of the asymptotic behavior of the dispersion curves for the relaxed micromorphic model with full inertia we refer to \cite{madeo2017role}.

\subsection{Energy flux in the 1D case}
When considering the conservation of total energy, it can be checked that
the first component of the energy flux (\ref{Flux}) can be rewritten
in terms of the new variables as

\begin{equation}
H_{1}=H_{1}^{1}+H_{1}^{2}+H_{1}^{3}+H_{1}^{4}+H_{1}^{5}+H_{1}^{6},\label{FluxRelaxed}
\end{equation}
with

\begin{flalign}
H_{1}^{1}= & \:{v}_{1,t}\cdot\left[\begin{pmatrix} -2\me-\lle & \hspace{-0.3cm}0 & 0\vspace{1.2mm}\\
0 & \hspace{-0.3cm}-\frac{\mLc}{2} & \mLc\vspace{1.2mm}\\
0 &\hspace{-0.3cm} \mLc & -2\,\mLc
\end{pmatrix}\cdot v_{1}'+\begin{pmatrix}-\frac{2\,\overline{\eta}_{1}+\overline{\eta}_{3}}{3} & 0 & 0\vspace{1.2mm}\\
0 &0 & 0\vspace{1.2mm}\\
0 & 0 & 0
\end{pmatrix}\cdot v_{1,tt}'+\begin{pmatrix}0 & 2\me & 2\me+3\lle\vspace{1.2mm}\\
0 & 0 & 0\vspace{1.2mm}\\
0 & 0 & 0
\end{pmatrix}\text{\ensuremath{\cdot}}v_{1}\right],\vspace{1.2mm}\nonumber \\
H_{1}^{2}= & \:{v}_{2,t}\cdot\left[\begin{pmatrix}-\left(\me+\mc\right) &  \hspace{-0.3cm}0 & 0\vspace{1.2mm}\\
0 & \hspace{-0.3cm} -\mLc & -\mLc\vspace{1.2mm}\\
0 & \hspace{-0.3cm} -\mLc & -\mLc
\end{pmatrix}\cdot v_{2}'+\begin{pmatrix}-\frac{\overline{\eta}_{1}+\overline{\eta}_{2}}{2} & 0 & 0\vspace{1.2mm}\\
0 &0 & 0\vspace{1.2mm}\\
0 & 0 & 0
\end{pmatrix}\cdot v_{2,tt}'+\begin{pmatrix}0 & 2\me & -2\text{\ensuremath{\mu}}_{c}\vspace{1.2mm}\\
0 & 0 & 0\vspace{1.2mm}\\
0 & 0 & 0
\end{pmatrix}\text{\ensuremath{\cdot}}v_{2}\right],\vspace{1.2mm}\label{FlLong}\\
H_{1}^{3}=\: & {v}_{3,t}\cdot\left[\begin{pmatrix}-\left(\me+\mc\right) & \hspace{-0.3cm} 0 & 0\vspace{1.2mm}\\
0 & \hspace{-0.3cm} -\mLc & -\mLc\vspace{1.2mm}\\
0 & \hspace{-0.3cm} -\mLc & -\mLc
\end{pmatrix} \cdot v_{3}'+\begin{pmatrix}-\frac{\overline{\eta}_{1}+\overline{\eta}_{2}}{2} & 0 & 0\vspace{1.2mm}\\
0 &0 & 0\vspace{1.2mm}\\
0 & 0 & 0
\end{pmatrix}\cdot v_{3,tt}'+\begin{pmatrix}0 & 2\me & -2\text{\ensuremath{\mu}}_{c}\vspace{1.2mm}\\
0 & 0 & 0\vspace{1.2mm}\\
0 & 0 & 0
\end{pmatrix}\text{\ensuremath{\cdot}}v_{3}\right],\vspace{1.2mm}\nonumber \\
H_{1}^{4}= & -2\mLc\left(v_{4}\right)_{,1}{v}_{4,t},\qquad H_{1}^{5}=-2\mLc\left(v_{5}\right)_{,1}{v}_{5,t},\qquad H_{1}^{6}=-\frac{\mLc}{2}\left(v_{6}\right)_{,1}{v}_{6,t}.\nonumber 
\end{flalign}

On the other hand, for Cauchy continua, the first component of the energy flux can be written as:
\begin{equation}
H_{1}=H_{1}^{1}+H_{1}^{2}+H_{1}^{3},\label{FluxCauchy1}
\end{equation}
where:
\begin{align}
H_{1}^{1}&=-(2\,\lambda+\mu) \,u_1'\, u_{1,t}, \qquad\qquad H_{1}^{2}=-\mu\, u_2'\, u_{2,t},\qquad\qquad H_{1}^{3}=-\mu\, u_2'\, u_{2,t}.
\end{align}

\section{Reflection and transmission  at a Cauchy/relaxed-micromorphic interface\label{Reflection}}

In this section, we present the results obtained with a specific choice of boundary conditions to be imposed between a Cauchy medium and a relaxed micromorphic medium. Such set of boundary conditions has been derived in \cite{madeo2016reflection}, used in \cite{madeo2016first} and it allows to describe free vibrations of the microstructure at the considered interface. For the full presentation of the complete sets of possible connections that can be established at Cauchy/relaxed, relaxed/relaxed, Cauchy/Mindlin, Mindlin/Mindlin interfaces we refer to\cite{madeo2016reflection}.

The main aim of this section is toextend the results presented in \cite{madeo2016reflection} to the case of a kinetic energy with free and gradient micro-inertia and to show the peculiar effect that the gradient micro-inertia may have on the reflection and transmission coefficients. 

As it has been shown in \cite{madeo2016modeling}, the role of the gradient micro-inertia is essential for the fitting of the relaxed model on real band-gap metamaterials.

\subsection{Interface jump conditions at a Cauchy/relaxed-micromorphic interface \label{Interface}}
When considering connections between a Cauchy and a relaxed micromorphic medium one can impose more kinematical boundary conditions than in the case of connections between Cauchy continua. More precisely, one can act on the displacement field $\text{\ensuremath{ u}}$ (on both sides of the interface) and also on the tangential part of the micro-distortion
$\text{\ensuremath{\p}}$ (on the side of the interface occupied
by the relaxed micromorphic continuum).  In what follows, we consider
the ``-'' region occupied by the Cauchy continuum and the ``+''
region occupied by the micromorphic continuum and we denote by $f$ and $t$ the forces  and by $\tau$  the double-forces, see \cite{madeo2016reflection}:
\begin{align}
f & =  \Sig^{-}\cdot \n^{-},\qquad  t=\left(\sig^{+}+\mathcal{I}^+\right)\cdot \n^{+},\qquad\tau= \mLc\,(	\Curl\,\p^{+}) \x \epsilon \x \n^{+}.
\end{align}
It is possible to check that considering the normal $\n=(1,0,0)$, the normal components $\tau_{11},\tau_{21}$ and $\tau_{31}$ of the double force are identically zero. Therefore, the number of independent conditions that one can impose on the micro-distortions is 6 when considering a relaxed micromorphic model.

In this paper we focus the attention on one particular type of connection between a classical Cauchy continuum and a relaxed micromorphic one, which is sensible to reproduce the real situation in which the microstructure of the band-gap metamaterial is free to vibrate independently of the macroscopic matrix. Such a particular connection guarantees the continuity of the macroscopic displacement and free motion of the microstructure (which means vanishing double force) at the interface (see \cite{madeo2016reflection}):
\begin{align}
\left[[ u\right]] =0,\qquad\qquad  t- f=0,\qquad\qquad\tau\cdot\nu_{2}=\tau\cdot\nu_{3}=0,
\end{align}
where, again, $\nu_{2}$ and $\nu_{3}$ are two independent unit normal vectors tangent to the interface.

We explicitly remark that the continuity of displacement implies the continuity of the internal forces and that the conditions on the arbitrariness of micro-motions is assured by imposing that the tangential part of the double force is vanishing. 

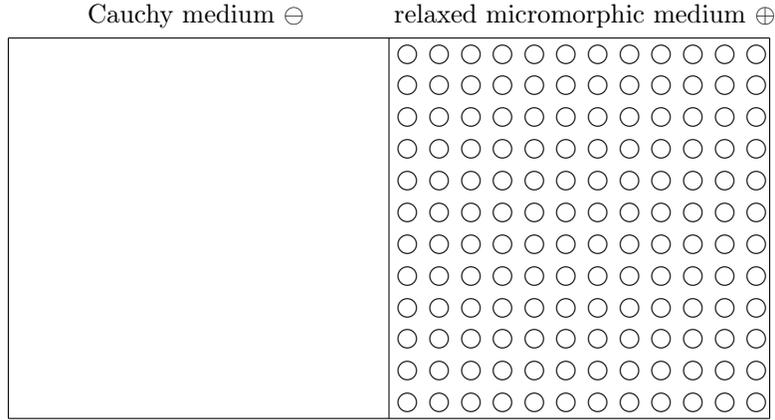
\begin{figure}[H]
	\begin{centering}
		
		\begin{picture}(288,160)
		\put(0,10){\line(0,1){144}}
		\put(0,10){\line(1,0){288}}
		\put(0,154){\line(1,0){288}}
		\put(144,10){\line(0,1){144}}
		\put(288,10){\line(0,1){144}}
		\multiput(151,16)(12,0){12}{
			\multiput(0,0)(0,12){12}{\circle{7.2}}}
		
		\put(30,160){Cauchy medium $\ominus$}		
		\put(146,160){relaxed micromorphic medium $\oplus$}
		

		\end{picture}
		\par\end{centering}
	
	\protect\caption{Schematics of a macro internal clamp with free microstructure at a Cauchy/relaxed-micromorphic interface \cite{madeo2016first,madeo2016reflection}.}
\end{figure}

Introducing the tangent vectors $\nu_{2}=(0,1,0)$ and $\nu_{3}=(0,0,1)$ and considering the new variables presented in \eqref{Variables}, the boundary conditions on the jump of displacement read:
\begin{align}
v_{1}^{+}\cdot\n-u_{1}^{-}=0,\qquad\qquad v_{2}^{+}\cdot\n-u_{2}^{-}=0,\qquad\qquad v_{3}^{+}\cdot\n-u_{3}^{-}=0,\label{Jump1}
\end{align}
while the conditions on the internal forces become (see also \cite{madeo2016reflection}):
\begin{align}
\left(\begin{array}{c}
\lle+2\me\\
0\\ 0 \end{array}\right)\cdot\left( v_{1}^{+}\right)'+\left(\begin{array}{c} \frac{2\,\overline{\eta}_{1}+\overline{\eta}_{3}}{3} 
\\
0\\ 0 \end{array}\right)\cdot\left( v_{1,tt}^{+}\right)'+\left(\begin{array}{c}
0\\ -2\me\\ -(3\lle+2\me) \end{array}\right)\cdot v_{1}^{+}&=(\lambda+2\mu)\,\left(u_{1}^{-}\right)',\nonumber 
\\\nonumber \\ 
\left(\begin{array}{c} \me+\mu_{c}\\ 0\\ 0 \end{array}\right)\cdot\left( v_{2}^{+}\right)'+\left(\begin{array}{c} \frac{\overline{\eta}_{1}+\overline{\eta}_{2}}{2} 
\\
0\\ 0 \end{array}\right)\cdot\left( v_{2,tt}^{+}\right)'+\left(\begin{array}{c} 0\\ -2\me\\ 2\mu_{c}
\end{array}\right)\cdot v_{2}^{+}&=\mu\,\left(u_{2}^{-}\right)',
\label{Jump2}\\\nonumber \\ 
\nonumber \left(\begin{array}{c} \me+\mu_{c}\\ 0\\ 0 \end{array}\right)\cdot\left( v_{3}^{+}\right)'+\left(\begin{array}{c} \frac{\overline{\eta}_{1}+\overline{\eta}_{2}}{2} 
\\
0\\ 0 \end{array}\right)\cdot\left( v_{3,tt}^{+}\right)'+\left(\begin{array}{c} 0\\ -2\me\\ 2\mu_{c} \end{array}\right)\cdot v_{3}^{+}&=\mu\,\left(u_{3}^{-}\right)'.
\end{align}
The conditions on the tangential part of the double force $\tau$  can be written as
\begin{align}
\tau_{22}&=\left(\begin{array}{c}
0\\ -\mLc/2\\ \mLc \end{array}\right)\cdot\left( v_{1}^{+}\right)'+\frac{\mLc}{2}\left(\mathrm{v}_{6}^{+}\right)'=0, 
&\tau_{33}&=\left(\begin{array}{c} 0\\ -\mLc/2\\ \mLc \end{array}\right)\cdot\left( v_{1}^{+}\right)'-\frac{\mLc}{2}\left(\mathrm{v}_{6}^{+}\right)'=0,
\nonumber \\\nonumber \\ 
\tau_{12}&=\left(\begin{array}{c} 0\\ \mLc\\ \mLc \end{array}\right)\cdot\left( v_{2}^{+}\right)'=0, &
\tau_{13}&=\left(\begin{array}{c} 0\\ \mLc\\ \mLc \end{array}\right)\cdot\left( v_{3}^{+}\right)'=0,
\label{Jump3}\\\nonumber \\ 
\tau_{23}&=\mLc\left(\left(\mathrm{v}_{4}^{+}\right)'+\left(\mathrm{v}_{5}^{+}\right)'\right)=0,
&\tau_{32}&=\mLc\left(\left(\mathrm{v}_{4}^{+}\right)'-\left(\mathrm{v}_{5}^{+}\right)'\right)=0,\nonumber 
\end{align}
while it can be verified that the normal part of the double force is identically vanishing, i.e.:
\begin{align}
\tau_{11}=0,\qquad\qquad
\tau_{21}=0,\qquad\qquad
\tau_{31}=0.
\end{align}

\subsection{Reflection and transmission coefficients at a Cauchy/relaxed-micromorphic interface}

We now want to define the reflection and transmission coefficients for the considered Cauchy/relaxed-micromorphic interface. To this purpose we introduce the quantities\footnote{We suppose here that $X=0$ is the position of the interface.}
\begin{align}
J_{i}=\int_{0}^{\Pi}H_{i}\left(0,t\right)dt,\qquad\qquad J_{r}=\int_{0}^{\Pi}H_{r}\left(0,t\right)dt,\qquad\qquad J_{t}=\int_{0}^{\Pi}H_{t}\left(0,t\right)dt, \label{J}
\end{align}
where $\Pi$ is the period of the traveling plane wave and $H_{i}\,$,
$H_{r}$ and $H_{t}$ are the energy fluxes of the incident, reflected
and transmitted energies, respectively. The reflection and transmission
coefficients can hence be defined as
\begin{equation}
R=\frac{J_{r}}{J_{i}},\qquad T=\frac{J_{t}}{J_{i}}.\label{eq:Reflection_coefficient}
\end{equation}
Since the considered system is conservative, one must have $R+T=1$.

The quantities $J_i$, $J_r$ and $J_t$ can be explicitly calculated once the fluxes $H_i$, $H_r$ and $H_t$ are computed for the particular type of connection presented in section \ref{Interface}. To compute such fluxes for the problem at hand, we set:
\begin{equation}
u_{1}^{-}(X,t)=u_{1}^{i}(X,t)+u_{1}^{r}(X,t),\quad\quad u_{2}^{-}(X,t)=u_{2}^{i}(X,t)+u_{2}^{r}(X,t),\quad\quad u_{3}^{-}(X,t)=u_{3}^{i}(X,t)+u_{3}^{r}(X,t),\nonumber
\end{equation}
where it can be found that for Cauchy media: 
\begin{align}
u_{1}^{i}(X,t)&=\bar{\alpha}_{1}\,e^{i(\frac{\omega}{c_{l}}\,X-\omega\,t)},
&u_{1}^{r}(X,t)&=\alpha_{1}\,e^{i(-\frac{\omega}{c_{l}}\,X-\omega\,t)},\vspace{1.2mm}\nonumber\\
u_{2}^{i}(X,t)&=\bar{\alpha}_{2}\,e^{i(\frac{\omega}{c_{t}}\,X-\omega\,t)},
&u_{2}^{r}(X,t)&=\alpha_{2}\,e^{i(-\frac{\omega}{c_{t}}\,X-\omega\,t)},\vspace{1.2mm}\label{CauchyProp}\\
u_{3}^{i}(X,t)&=\bar{\alpha}_{3}\,e^{i(\frac{\omega}{c_{t}}\,X-\omega\,t)},
&u_{3}^{r}(X,t)&=\alpha_{3}\,e^{i(-\frac{\omega}{c_{t}}\,X-\omega\,t)},\nonumber
\end{align}
with $c_l=\sqrt{\frac{\lambda+2\,\mu}{\rho}}$ and $c_t=\sqrt{\frac{\mu}{\rho}}$.

Moreover, solving the eigenvalue problems \eqref{Dispersion} (see also \cite{madeo2016reflection}), we find:
\begin{align}
v_{1}^{i}(X,t)&=\beta^{1}_{1}\, h^{1}_{1} \,e^{i(k^{1}_{1}\,X-\omega\,t)}+\beta^{2}_{1}\, h^{2}_{1} \,e^{i(k^{2}_{1}\,X-\omega\,t)},\qquad &v_{\alpha}^{i}(X,t)&=\beta^{1}_{\alpha}\, h^{1}_{\alpha} \,e^{i(k^{1}_{1}\,X-\omega\,t)}+\beta^{2}_{\alpha}\, h^{2}_{\alpha} \,e^{i(k^{2}_{\alpha}\,X-\omega\,t)},\vspace{1.2mm}\nonumber\\
v_{4}^{i}(X,t)&=\beta_{4}\,e^{i(\sqrt{\omega-\omega_s^2}\,X/c_{m1}-\omega\,t)}, \qquad &
v_{5}^{i}(X,t)&=\beta_{5}\,e^{i(\sqrt{\omega-\omega_s^2}\,X/c_{m2}-\omega\,t)},\vspace{1.2mm}\label{Modes}\\
v_{6}^{i}(X,t)&=\beta_{6}\,e^{i(\sqrt{\omega-\omega_s^2}\,X/c_{m1}-\omega\,t)},\nonumber
\end{align}
where $\pm k_1^1(\omega)$ and $\pm k_1^2(\omega)$ are the solutions of the first of Equations \eqref{Dispersion} \footnote{It can be checked that $\det (A_1)=0$ is a polynomial of the fourth order in $k$ involving only even powers of k. The sign + in Equation \eqref{Modes} is chosen because the transmitted wave has the same direction as the incident wave.} and $h_1^1(\omega)$ and $h_1^2(\omega)$ are the eigenvectors of the corresponding problem $A_1 \cdot \beta=0$. Analogously, $\pm k_\alpha^1(\omega)$ and $\pm k_\alpha^2(\omega)$ ($\alpha=2,3$) are the solutions of the second of Equations \eqref{Dispersion} and  $h_\alpha^1(\omega)$ and $h_\alpha^2(\omega)$ the corresponding eigenvectors.

With these notations, the flux associated to the incident, reflected and transmitted waves can be computed recalling Equations \eqref{Modes} and \eqref{Flux} as:
\begin{align}
H_i&=-(\lambda+2\,\mu)\,u^i_{1,t} \,u^i_{1,1}-\mu\,u^i_{2,t} \,u^i_{2,1}-\mu\,u^i_{3,t} \,u^i_{3,1}. \nonumber\\
H_r&=-(\lambda+2\,\mu)\,u^r_{1,t} \,u^r_{1,1}-\mu\,u^r_{2,t} \,u^r_{2,1}-\mu\,u^r_{3,t} \,u^r_{3,1}, \label{flux}\\
H_t&=H_1^1+H_1^2+H_1^3+H_1^4+H_1^5+H_1^6, \nonumber
\end{align}
where the transmitted flux is computed using the wave form solutions \eqref{Modes} in Equations \eqref{FlLong}. In this way, the only unknowns appearing in Equations \eqref{flux} are the 12 scalar amplitudes\footnote{The amplitudes of the incident waves $\bar{\alpha}_1$, $\bar{\alpha}_2$ and $\bar{\alpha}_3$ are supposed to be known.} $\alpha_1$, $\alpha_2$, $\alpha_3$, $\beta_1^1$, $\beta_1^2$, $\beta_2^1$, $\beta_2^2$, $\beta_3^1$, $\beta_3^2$, $\beta_3^3$, $\beta_4$, $\beta_5$, $\beta_6$ which can now be computed using the 12 boundary conditions \eqref{Jump1}, \eqref{Jump2} and \eqref{Jump3}.

\begin{figure}[H]
	\begin{centering}
		\includegraphics[width=10cm]{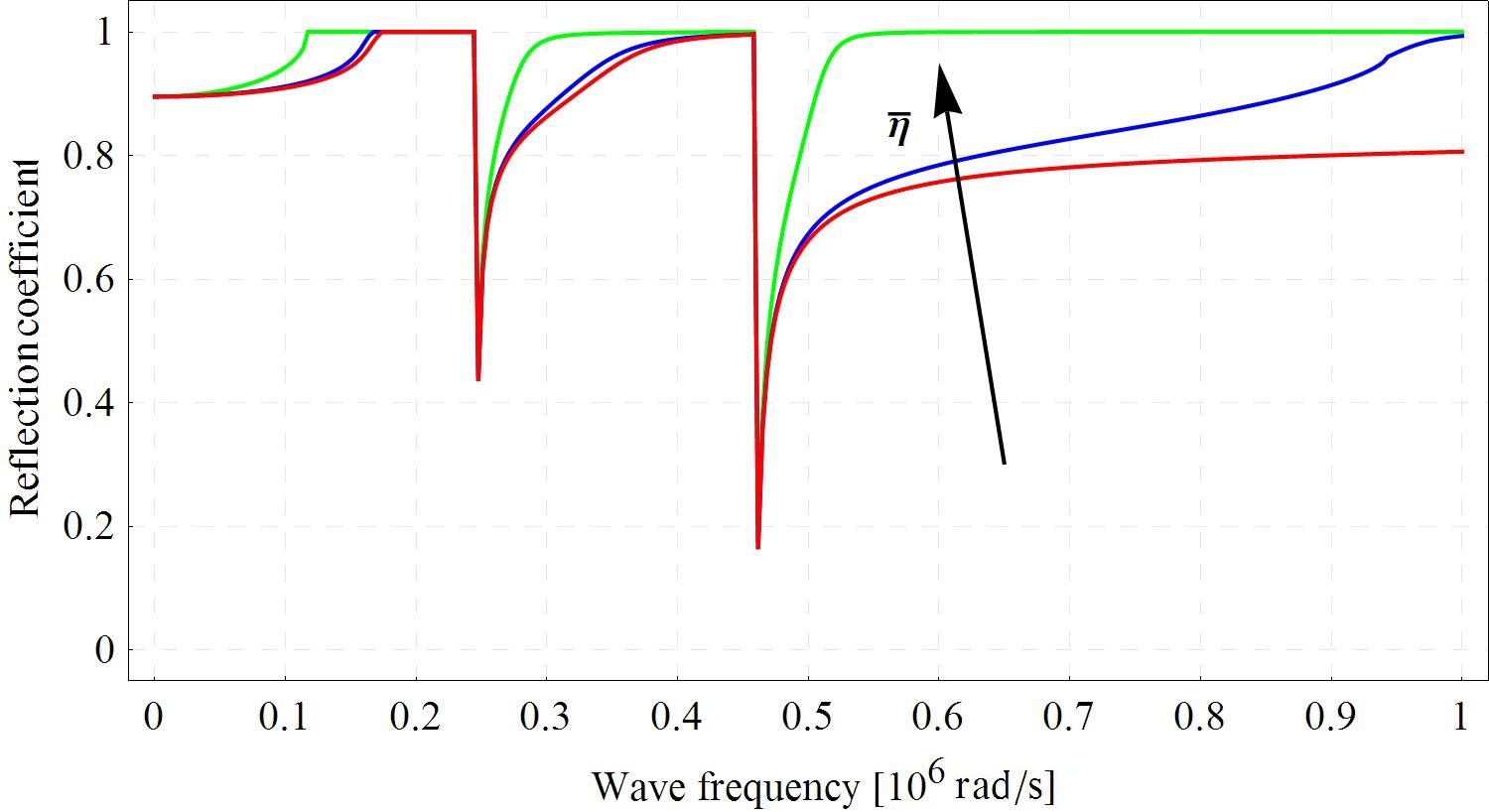} 
		\par\end{centering}
	\protect\caption{\label{fig:RP}\textcolor{black}{Cauchy/relaxed-micromorphic interface:
			macro clamp with free microstructure. }Reflection coefficient as
		function of frequency for incident longitudinal waves
		s with
		free micro-inertia $\eta_1=\eta_2=\eta_3=10^{-2}$ and gradient micro-inertia $\overline{\eta}_{3}=(3\times10^{-4},3\times10^{-3},3\times10^{-2})kg/m$.}
\end{figure}

\begin{figure}[H]
	\begin{centering}
		\includegraphics[width=10cm]{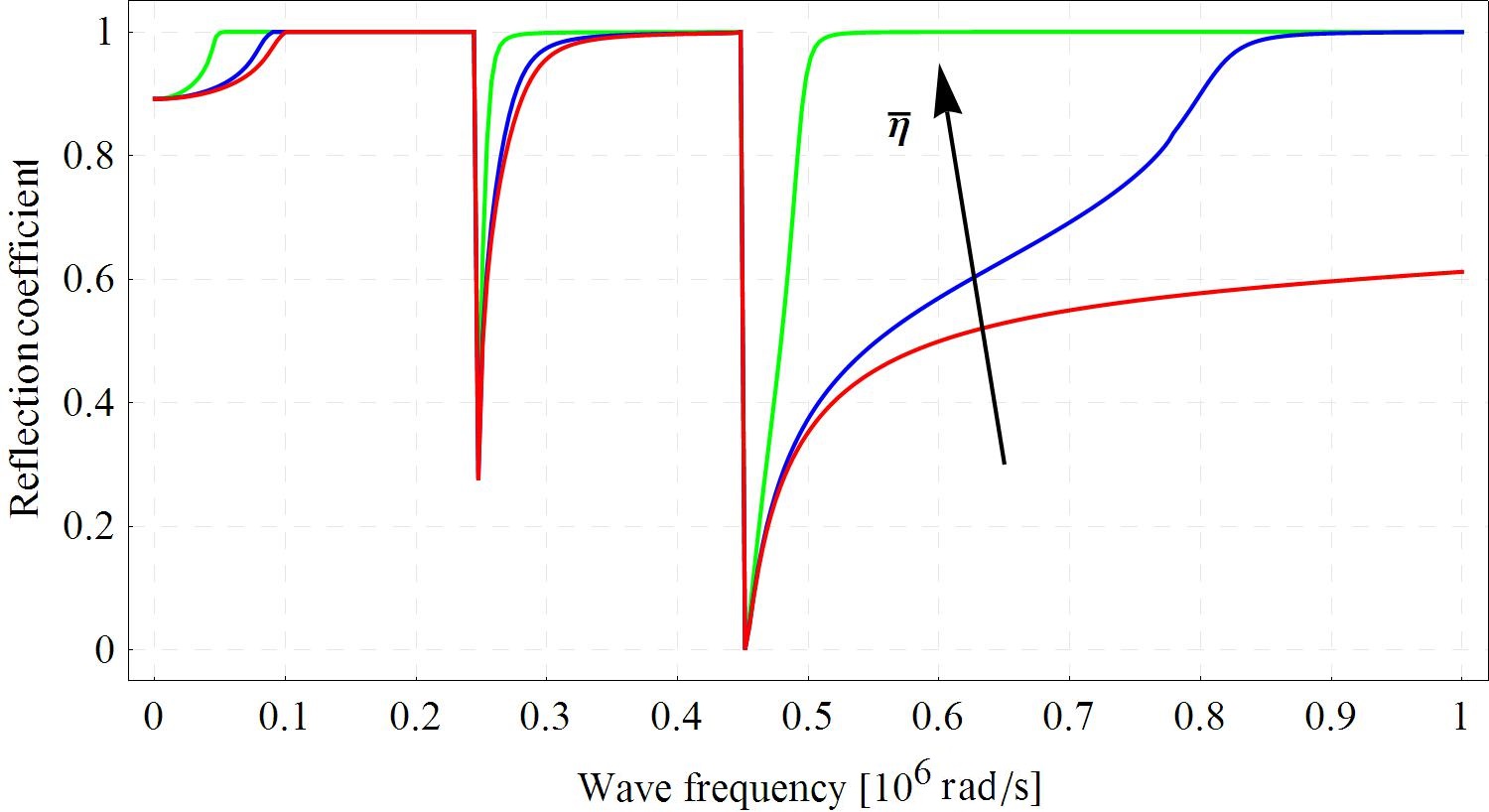} 
		\par\end{centering}
	\protect\caption{\label{fig:RS}\textcolor{black}{Cauchy/relaxed-micromorphic interface:
			macro clamp with free microstructure. }Reflection coefficient as
		function of frequency for incident transverse waves with micro-inertia $\eta_1=\eta_2=\eta_3=10^{-2}$ and
		gradient micro-inertia $\overline{\eta}_{2}=(2\times10^{-4},2\times10^{-3},2\times10^{-2})kg/m$.}
\end{figure}

\begin{figure}[H]
	\begin{centering}
		\includegraphics[width=10cm]{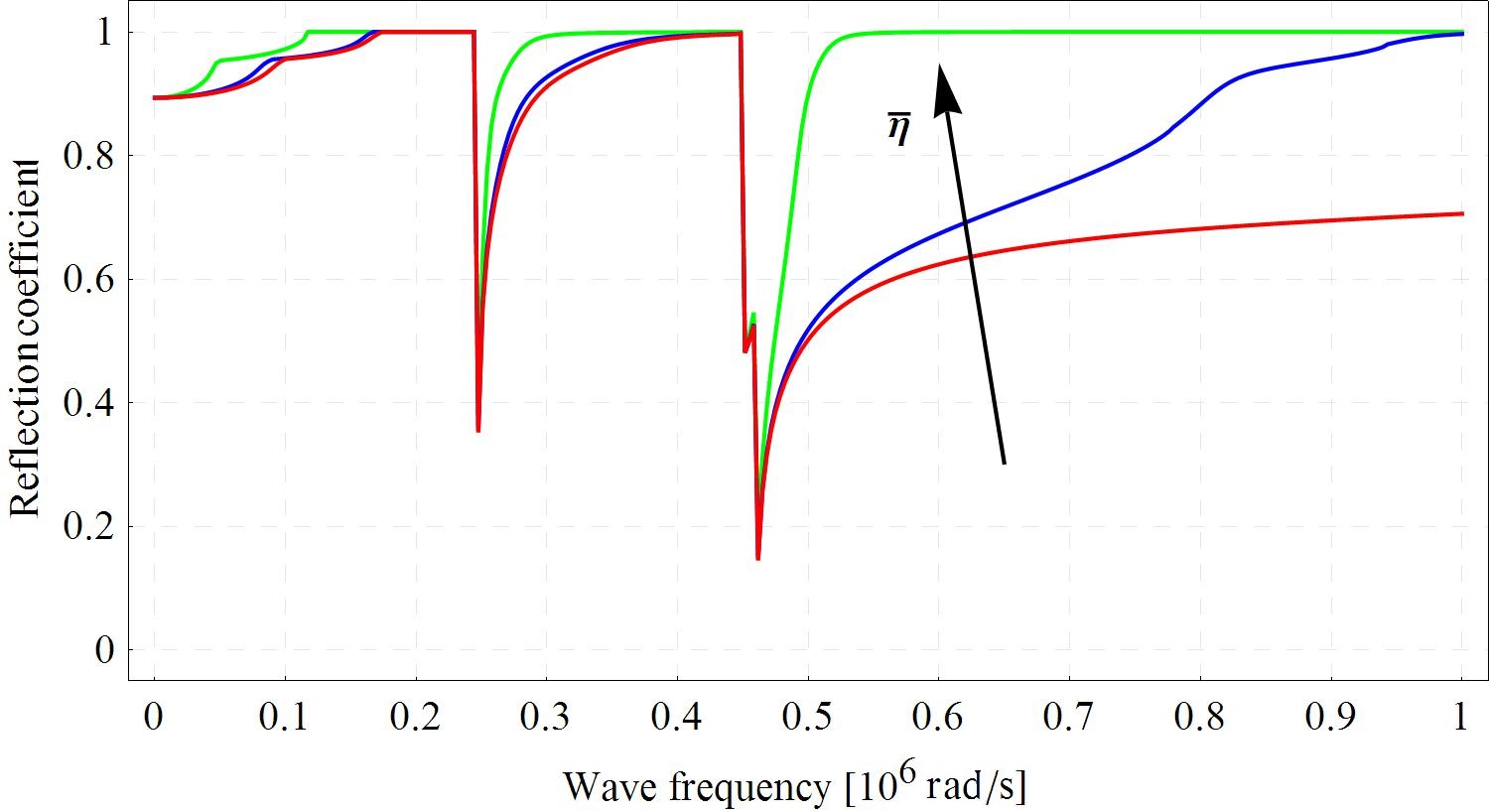} 
		\par\end{centering}
		\protect\caption{\label{fig:RPS}\textcolor{black}{Cauchy/relaxed-micromorphic interface:
				macro clamp with free microstructure. }Reflection coefficient as
			function of frequency for an incident wave with both longitudinal and transverse components with
			free micro-inertia $\eta_1=\eta_2=\eta_3=10^{-2}$, and gradient micro-inertiae $\overline{\eta}_{3}=(3\times10^{-4},3\times10^{-3},3\times10^{-2})kg/m$ and $\overline{\eta}_{2}=(2\times10^{-4},2\times10^{-3},2\times10^{-2})kg/m$.}	
\end{figure}

Once the amplitudes (and hence the solution) have been determined, the incident, reflected and transmitted flux can be computed according to equation \eqref{flux}. The reflection and transmission coefficients $R$ and $T$ can be computed according to Eqs. \eqref{eq:Reflection_coefficient}.  In Figures \ref{fig:RP}, \ref{fig:RS}, we show the behavior of the reflection coefficient $R$ as a function of frequency for longitudinal and transverse waves separately, while in  \ref{fig:RPS}, we show the result obtained with a combination of both.

Since the case of pure longitudinal and pure transverse incident wave is completely analogous, we only comment here the behavior of the reflection coefficient for the combined waves, with reference to Figure \ref{fig:RPS}. Firstly, we describe the characteristics of the reflection spectrum that are independent of the effects of the gradient micro-inertia. It can be seen that a first frequency interval, for which complete reflection occurs, can be observed. This interval corresponds to the band-gap observed analyzing only the bulk propagation (dispersion curves). A second band-gap can be observed for higher frequencies which is uniquely due to the presence of the interface. Furthermore, some phenomena of localized resonance occur around $\omega=0.2\times10^{6}\,$rad/s and $\omega=0.4\times10^{6}\,$rad/s for both longitudinal and transverse waves. These effects have to be related to the fact that the cut-off frequencies $\omega_{p}$ and $\omega_{r}$ are indeed resonance frequencies for the considered free microstructure. Such peaks of reflected energy can hence be completely associated to the characteristics of the considered microstructures and to their characteristic resonant behaviors. 

The presented effects are specific of the relaxed micromorphic model (even without gradient micro-inertia) and have already been studied in \cite{madeo2016reflection}, and applied in \cite{madeo2016first} to  the modeling of real phononic crystals of the type studied in \cite{lucklum2012two}. In this paper we focus on the addition of a gradient micro-inertia and on the consequent effects on the reflection spectra. It is possible to notice that, when the value of the gradient micro-inertia is sufficiently high a third band gap shows up. Moreover, the first and second  band-gap are extended. The resulting reflection spectrum shows transmission only for low values of frequency and for two additional values of frequency (internal resonances). Moreover, no transmission at all is allowed for higher frequencies.

\section{Modeling the transmission spectra of a specific microstructure with a FEM model and a relaxed micromorphic continuum\label{FEM}}

In this section we apply the results presented in the previous sections to describe the reflective behavior of the interface between an aluminum plate (modeled as a classical Cauchy continuum) and a metamaterial with specific microstructure (modeled via a relaxed micromorphic model). To show the validity of the enriched continuum modeling framework previously introduced we will compare it with direct FEM simulations performed with the code COMSOL\textsuperscript{\textregistered} Multiphysics.

\subsection{Microstructure and material parameters}
\begin{figure}[H]
	\begin{centering}
\begin{picture}(410,131)	
	\put(7,0){\includegraphics[height=5cm]{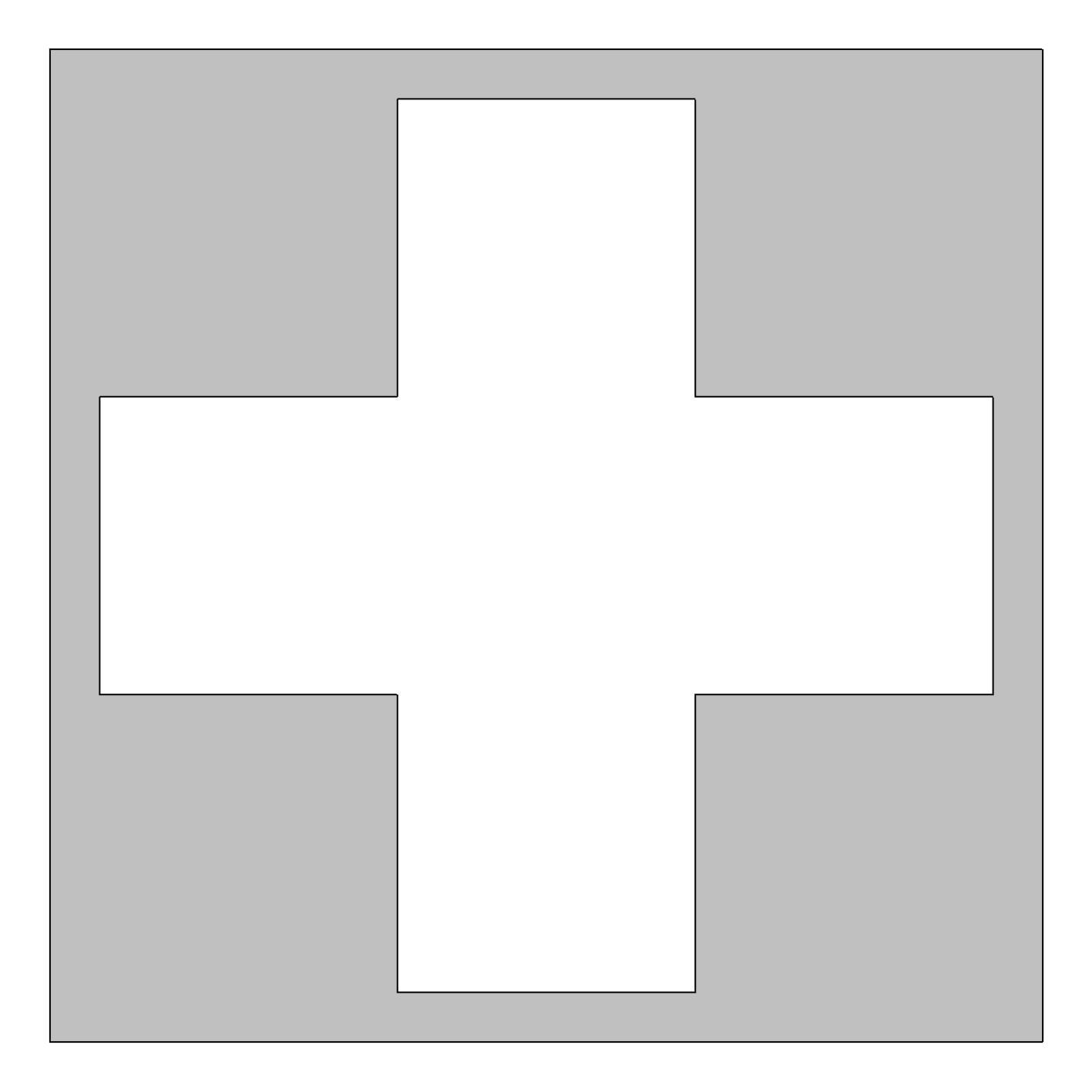}}
	\put(2,6){\line(0,1){130}}
	\put(0,6){\line(1,0){4}}
	\put(0,136){\line(1,0){4}}
	\put(4,70){a}
	\put(78,13){\line(0,1){116.5}}
	\put(76,13){\line(1,0){4}}
	\put(76,129.5){\line(1,0){4}}
	\put(82,70){b}
	\put(28,51.5){\line(0,1){39}}
	\put(26,51.5){\line(1,0){4}}
	\put(26,90.5){\line(1,0){4}}
	\put(30,70){c}
	\put(170,80){	\begin{tabular}{|c|c|c|c|c|c|}
	
		\hline 
		a & b & c & $\rho$ & $E$ & $\nu$\tabularnewline
		\hline 
		{[}$\mathrm{mm}${]} & {[}$\mathrm{mm}${]} & {[}$\mathrm{mm}${]} & {[}$\mathrm{kg/m^{3}}${]} & {[}$\mathrm{GPa}${]} & {[}$-${]}\tabularnewline
		\hline 
		\hline 
		$1$ & $0.9$ & $0.3$ & $2700$ & $70$ & $0.33$\tabularnewline
		\hline 
		\end{tabular}}
	
\end{picture}

		\par\end{centering}
	\caption{Microstructure of the considered  metamaterial (left), values of the elastic parameters of the base material (aluminum) and geometric parameters relative to the unit cell (right). \label{fig:Microstructure}}
	
\end{figure}
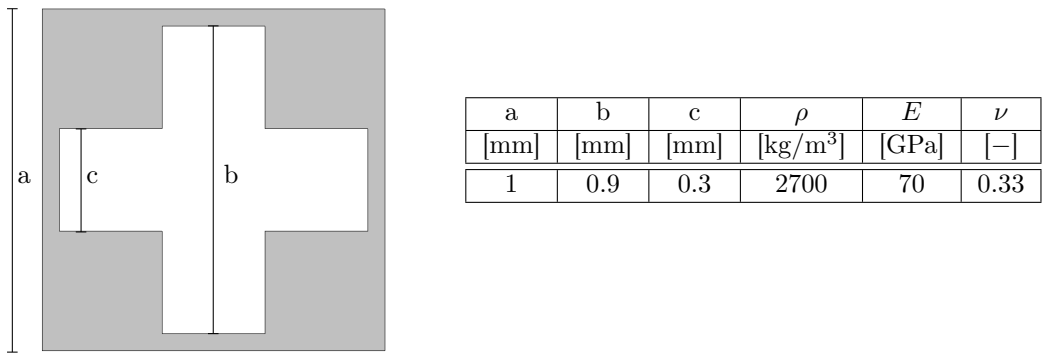

\begin{figure}[H]
	\begin{centering}
			\begin{picture}(550,180)
			\scalebox{0.87}{
				\put(345,0){	\includegraphics[width=6.5cm]{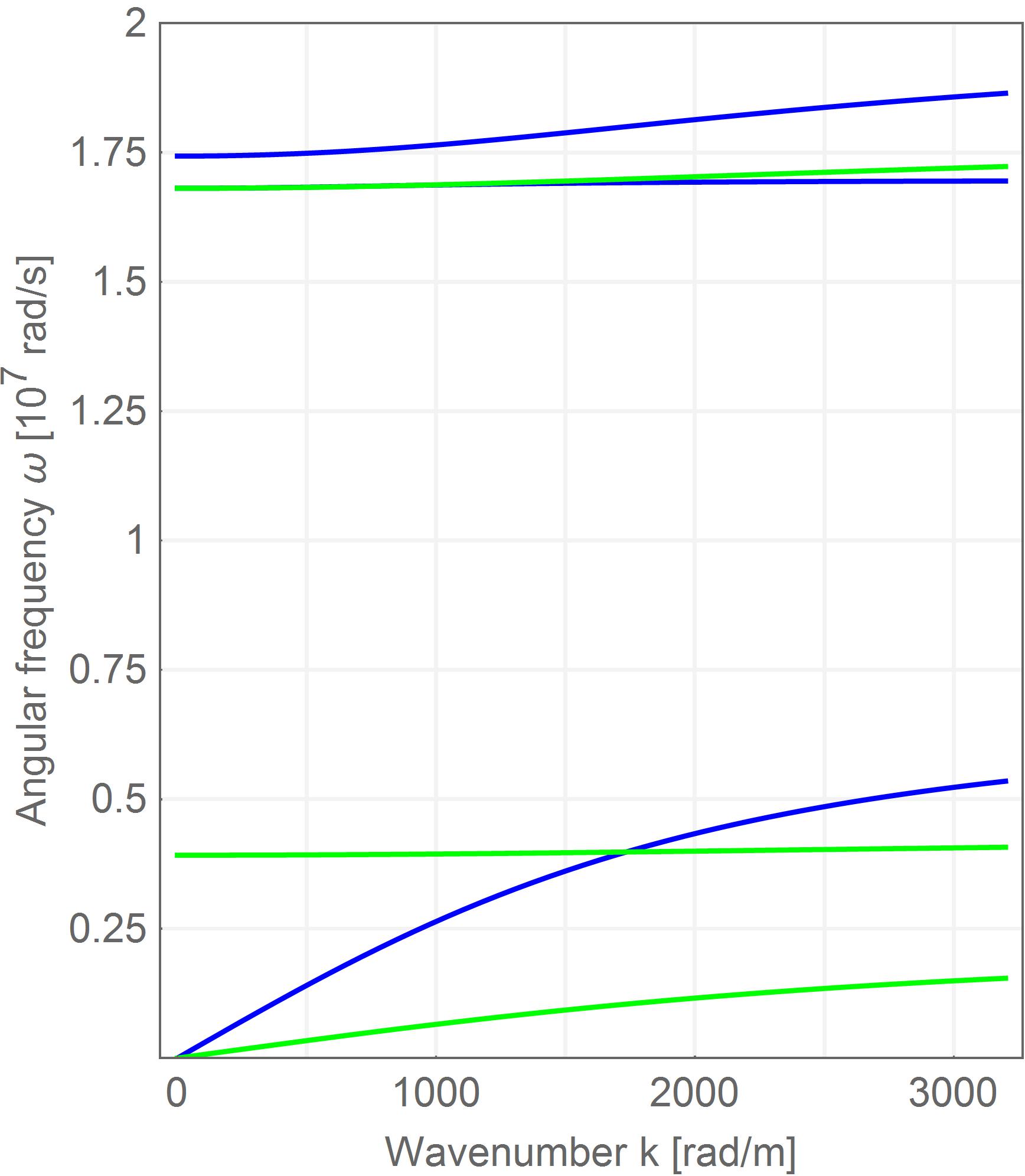}	}
						\put(170,0){	\includegraphics[width=6.5cm]{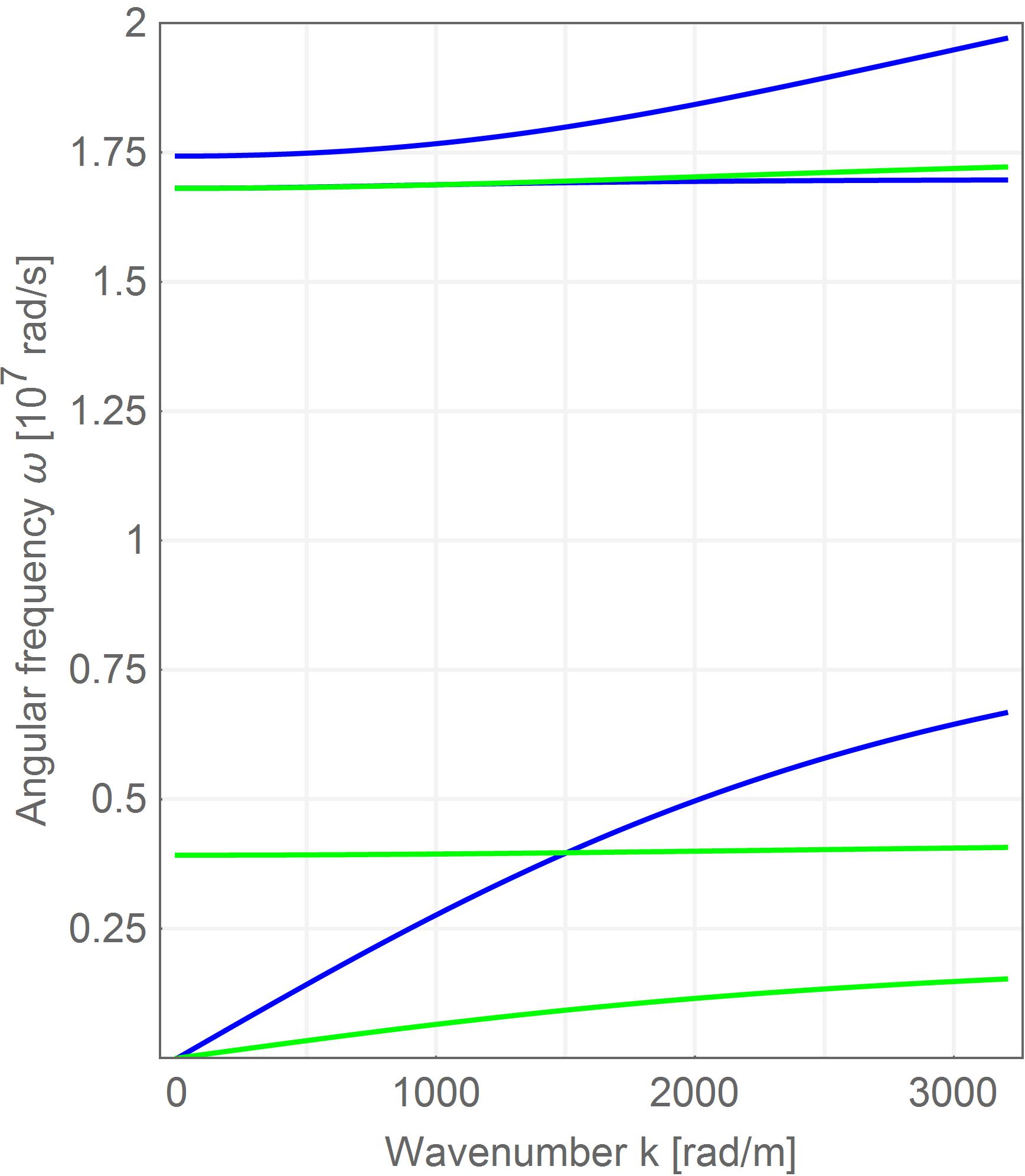}}
			\put(0,6){\includegraphics[width=6.5cm]{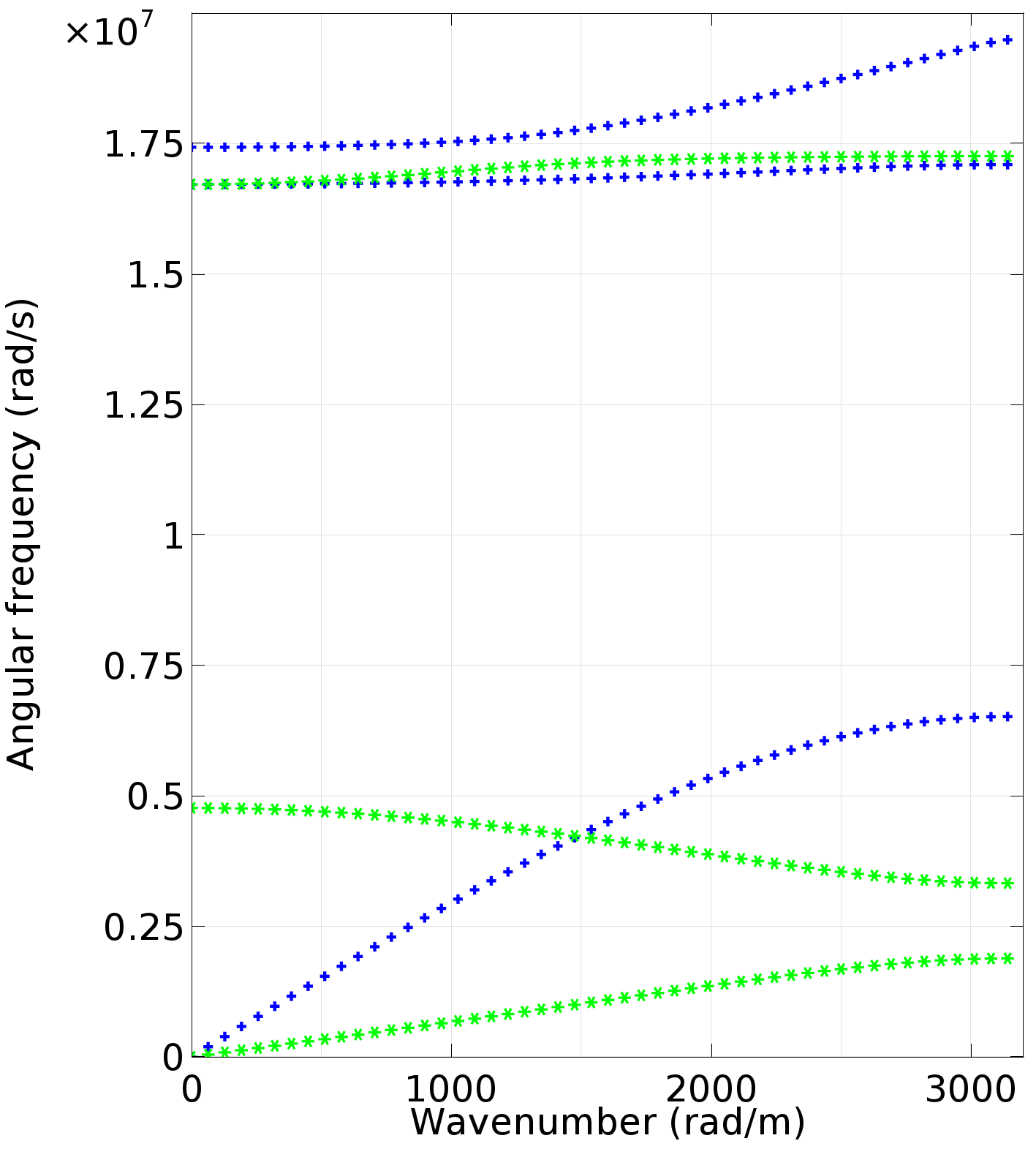}}
		}	
			\end{picture}
	 
		\par\end{centering}
	\protect\caption{\label{fig:Dispersion}Dispersion curves obtained with a Bloch wave analysis of the cell shown in Figure \ref{fig:Microstructure} (left) and dispersion curves for the relaxed micromorphic model with the parameters measured in \cite{madeo2016modeling} (center) and with the slightly modified set of the relaxed parameters value proposed in Table \ref{tab:parameters} (right). In blue are represented the longitudinal modes while in green the transverse ones.}	
\end{figure}

In a previous work \cite{madeo2016modeling}, we showed that the analysis of bulk wave propagation in a  metamaterial with periodic cross-like holes (see Figure  \ref{fig:Microstructure}) can be achieved using the relaxed micromorphic model. More particularly, by using a suitable fitting procedure, we have been able to calibrate the parameters of the relaxed model by superimposing the dispersion curves obtained via the relaxed model to those obtained via a Bloch wave analysis of the microstructure shown in Figure \ref{fig:Microstructure}. The results of the fitting procedure proposed in \cite{madeo2016modeling} are recalled in Figure \ref{fig:Dispersion} in which the dispersion curves obtained via the relaxed micromorphic model are compared to those issued via a Bloch wave analysis. In the same Figure, we also propose a slight variation of the parameters fitted in \cite{madeo2016modeling} which may provide a more precise result when considering the transmission spectra. All the material parameters considered are given in Table \ref{tab:parameters}.   The objective here is to show that the parameters derived in \cite{madeo2016modeling} using the bulk dispersion curves alone are true material constants that allow to describe the mechanical behavior of the considered metamaterial even when considering a complex (meta-) structure of the type presented in Figure  \ref{fig:Microstructure}. In particular, we will show that such parameters properly describe the mechanical behavior of the chosen metamaterial so well that the behavior of that metamaterial can be successfully described in more complex situations as the reflection and transmission at discontinuity interfaces of the material properties. The interest of using an enriched continuum model  resides in the unique possibility it offers to exploit only few material parameters  for the description of the mechanical behavior of an otherwise rather complicated system.

In \cite{madeo2016modeling}, the material parameters of the equivalent relaxed micromorphic model were also determined. The values are shown in Table \ref{tab:parameters}.

\begin{table}[H]
	\begin{centering}
					\begin{tabular}{|c|c|c|c|c|c|c|}
			\hline 
			$\rho$  & $\mu_{c}$ & $\lambda_{\mathrm{micro}}$ & $\mu_{\mathrm{micro}}$ & $\lambda_{e}$ & $\mu_{e}$ & $L_{c}$ \tabularnewline
			\hline 
			$\left[\mathrm{kg/m^{3}}\right]$ & $\left[\mathrm{GPa}\right]$ & $\left[\mathrm{GPa}\right]$ & $\left[\mathrm{GPa}\right]$ & $\left[\mathrm{GPa}\right]$ & $\left[\mathrm{GPa}\right]$ & $\left[\mathrm{m}\right]$\tabularnewline
			\hline 
			$1323$ & $0.272$ & $19.8$ & $0.737$ & $17.7$ & $3.857$ & $0$\tabularnewline
			\hline 
		\end{tabular}
	\\	\vspace{0.3cm}
		\begin{tabular}{|c|c|c|c|c|c|c|c|}
			\hline 
			$\eta_{1}$  & $\eta_{2}$  & $\eta_{3}$  & $\bar{\eta}_{1}$  & $\bar{\eta}_{2}$ & $\bar{\eta}_{3}$& $\bar{\eta}_{2}$  from \cite{madeo2016modeling}   & $\bar{\eta}_{3}$  from \cite{madeo2016modeling}  \tabularnewline
			\hline 
			$\left[\mathrm{kg/m}\right]$ & $\left[\mathrm{kg/m}\right]$ & $\left[\mathrm{kg/m}\right]$ & $\left[\mathrm{kg/m}\right]$ & $\left[\mathrm{kg/m}\right]$ & $\left[\mathrm{kg/m}\right]$ & $\left[\mathrm{kg/m}\right]$ & $\left[\mathrm{kg/m}\right]$\tabularnewline
			\hline 
			$3.25\times10^{-5}$ & $3.25\times10^{-5}$ & $4\times10^{-4}$ & $0$ & $2\times10^{-4}$ & $6\times10^{-4}$ & $0.3\times10^{-4}$ & $1.8\times10^{-4}$\tabularnewline
			\hline 		
		\end{tabular}
\par\end{centering}
	\caption{\label{tab:parameters}Values of the material parameters (top) and of the micro-inertia parameters (bottom) of the weighted relaxed micromorphic model. All parameters are the same as the ones measured in \cite{madeo2016modeling} except for $\bar{\eta}_{2}$ and $\bar{\eta}_{3}$.}
\end{table}


For a comprehensive description of Figure \ref{fig:Dispersion}, the reader is referred to \cite{madeo2016modeling}. Here, we limit ourselves to point out the very good description of the band-gap and of the general behavior. The only difference between the two  approaches is given by the absence of a decreasing behavior in the first transverse optic mode. However, the average behavior of that vibrational mode is still well described. 

\subsection{FEM model for the determination of the transmission spectra \label{Spectra}}

\begin{figure}[H]
	\begin{centering}
		\scalebox{1.1}{		
			\begin{picture}(405,55)
			\put(15,23){
				\scalebox{0.7}{				
					\put(-18,-1){\vector(1,0){15}}
					\put(-18,4){\vector(1,0){15}}}
				
				\put(0,0){\line(1,0){390}}
				\put(0,3){\line(1,0){390}}
				\put(0,-10){\line(1,0){390}}
				\put(0,-12){\line(0,1){4}}
				\put(180,-12){\line(0,1){4}}
				\put(300,-12){\line(0,1){4}}
				\put(390,-12){\line(0,1){4}}

				
				\multiput(180,0)(3,0){40}{
					\put(0.15,0.95){\line(0,1){0.9}}
					\put(0.15,0.95){\line(1,0){0.9}}
					\put(2.85,0.95){\line(0,1){0.9}}
					\put(2.85,0.95){\line(-1,0){0.9}}
					\put(0.95,0.15){\line(1,0){0.9}}
					\put(0.95,0.15){\line(0,1){0.9}}
					\put(0.95,2.85){\line(1,0){0.9}}
					\put(0.95,2.85){\line(0,-1){0.9}}
					\put(0.15,1.85){\line(1,0){0.9}}
					\put(2.85,1.85){\line(-1,0){0.9}}
					\put(1.85,0.15){\line(0,1){0.9}}
					\put(1.85,2.85){\line(0,-1){0.9}}
				}
				
				\multiput(300,0)(0.6,0){150}{\line(0,1){3}}
				
				\put(390,0){\line(0,1){3}}
				
				\put(-15,30){excitation boundary}

				\put(0,-2){\thicklines\line(0,1){7}}	
				\put(-12,8){X=0}
				
				\put(90,-2){\thicklines\line(0,1){7}}	
				\put(78,8){X=$\overline{X}$}	
				
				\put(90,30){periodic boundary conditions}
				
				\put(-2,28){\scalebox{0.7}{	\vector(0,-1){13}}}
				
				\put(135,28.5){\scalebox{0.7}{	\vector(0,-1){41}}}
				\put(110,28.5){\scalebox{0.7}{	\vector(0,-1){36.5}}}
				
			}

			\put(95,0){Cauchy}		
			\put(205,0){microstructured material}
			\put(350,0){PML}	
			
			\end{picture}
		}
		
		\par\end{centering}
	
	\protect\caption{Schematic representation of the FEM model for the determination of the transmission spectrum, as implemented in COMSOL\textsuperscript{\textregistered}. \label{fig:geometry}}
\end{figure}
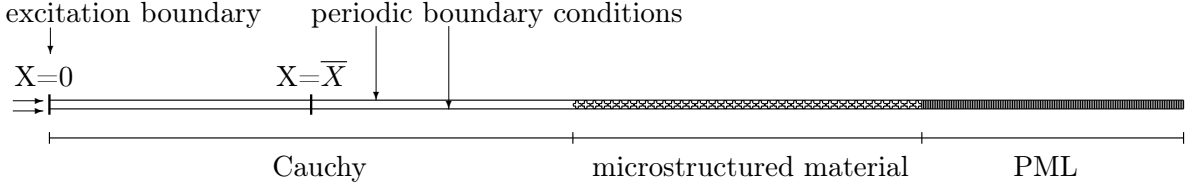

The transmission spectra for an  Cauchy-material/metamaterial interface is determined via FEM according to the model represented in Figure \ref{fig:geometry}. An external excitation is applied on the left side by imposing a unitary harmonic displacement and an analysis for different frequencies is performed by using the structural package of COMSOL\textsuperscript{\textregistered} Multiphysics. The incident wave propagates in the first part of the geometry that consists in an homogeneous strip of the Cauchy material. After a length of 0.6 m an array of 40 unit cells with cross-like holes of the type described in Figure \ref{fig:Microstructure} is added. When the wave arrives at the interface, it is partially reflected and partially transmitted. Finally, a Perfectly Matched Layer (PML) is added at the end of the strip to dissipate the transmitted wave and avoid spurious reflections in the metamaterial. On the upper and lower boundary, a periodic condition is applied to impose the propagation along the strip direction, thus reproducing the condition of plane wave propagation. The thickness of the strip is set to be equal to the height of the unit cell.

We are considering an interface between a classical Cauchy continuum on the $-$ side and a microstructured material on the $+$ side. The object is to see how much energy is transmitted through the interface. A priori, we have no information about the wave propagation in the microstructured material but the propagation in the homogeneous Cauchy continuum, if far enough from the interface, can be expected to take the usual form of waves propagating in Cauchy media. Therefore, we expect to find longitudinal waves propagating with wavelengths $k_l=\pm\frac{\omega}{c_l}$ and transverse waves with $k_t=\pm\frac{\omega}{c_t}$, where $c_l=\sqrt{\frac{\lambda+2\,\mu}{\rho}}$ and $c_t=\sqrt{\frac{\mu}{\rho}}$. The plus or minus sign in the wavenumber is due to the possibility that the waves can travel in both directions. Thus, we have that the waveform solution in the Cauchy material can be written as (see also Equation \eqref{CauchyProp}):
\begin{equation}
u_{1}^{-}(X,t)=u_{1}^{i}(X,t)+u_{1}^{r}(X,t),\quad\quad u_{2}^{-}(X,t)=u_{2}^{i}(X,t)+u_{2}^{r}(X,t),\quad\quad u_{3}^{-}(X,t)=u_{3}^{i}(X,t)+u_{3}^{r}(X,t),\nonumber
\end{equation}
where, as before, we set: 
\begin{align}
u_{1}^{i}(X,t)&=\bar{\alpha}_{1}\,e^{i(\frac{\omega}{c_{l}}\,X-\omega\,t)},
&u_{1}^{r}(X,t)&=\alpha_{1}\,e^{i(-\frac{\omega}{c_{l}}\,X-\omega\,t)},\vspace{1.2mm}\nonumber\\
u_{2}^{i}(X,t)&=\bar{\alpha}_{2}\,e^{i(\frac{\omega}{c_{t}}\,X-\omega\,t)},
&u_{2}^{r}(X,t)&=\alpha_{2}\,e^{i(-\frac{\omega}{c_{t}}\,X-\omega\,t)},\vspace{1.2mm}\label{CauchyProp2}\\
u_{3}^{i}(X,t)&=\bar{\alpha}_{3}\,e^{i(\frac{\omega}{c_{t}}\,X-\omega\,t)},
&u_{3}^{r}(X,t)&=\alpha_{3}\,e^{i(-\frac{\omega}{c_{t}}\,X-\omega\,t)}.\nonumber
\end{align}
Given the frequency $\omega$ of the traveling wave, the solution is hence known except for the 6 amplitudes $\bar{\alpha}_{i}$ and $\alpha_{i}$ ($i=1,2,3$).

In order to evaluate the unknown amplitudes in Equations \eqref{CauchyProp2} we can use the direct solution obtained via the FEM simulation. In particular, the solution for the displacement field obtained via the FEM code can be interpreted as the solution at a given instant (e.g. $t=0$) for every point of the domain. Considering the points $X=0$ and $X=\overline{X}$  (see Figure \ref{fig:geometry}) and the instant $t=0$, we can set a system of equations to compute the unknown amplitudes, in formulas:
 \begin{align}
 u_{1}^{-}(0,0)&=\bar{\alpha}_{1}+\alpha_1, &u_{2}^{-}(0,0)&=\bar{\alpha}_{2}+\alpha_2,
 &u_{3}^{-}(0,0)&=\bar{\alpha}_{3}+\alpha_3,\label{conditions}\\\nonumber
  u_{1}^{-}(\overline{X},0)&=\bar{\alpha}_{1}\,e^{i\,\frac{\omega}{c_{l}}\,\overline{X}}+\alpha_1\,e^{-i\,\frac{\omega}{c_{l}}\,\overline{X}},
 &u_{2}^{-}(\overline{X},0)&=\bar{\alpha}_{2}\,e^{i\,\frac{\omega}{c_{t}}\,\overline{X}}+\alpha_2\,e^{-i\,\frac{\omega}{c_{t}}\,\overline{X}}, &u_{3}^{-}(\overline{X},0)&=\bar{\alpha}_{3}\,e^{i\,\frac{\omega}{c_{t}}\,\overline{X}}+\alpha_3\,e^{-i\,\frac{\omega}{c_{t}}\,\overline{X}}.
 \end{align}
 where $ u_{i}^{-}(0,0)$ and $u_{i}^{-}(\overline{X},0)$ are known from the results of the FEM simulation.
 
From this system, it is possible to evaluate the unknown amplitudes and, therefore, the incident and reflected flux by using the first and second equations in \eqref{flux}. Finally, the reflection and transmission coefficients can be computed as:
\begin{equation}
R=\frac{J_{r}}{J_{i}},\qquad \qquad T=1-R.
\end{equation} 
We note that this semi-analytical procedure is valid only if the propagation of waves in the Cauchy material is planar in the FEM solution. It can happen, usually at high frequencies, that the resulting vibrational mode does not respect this assumption in which case the transmission spectra obtained applying this method may not be completely correct. However, in the range of frequencies considered here, the solution is constant along the section and the waveform evaluated with the resulting amplitudes is perfectly described by the solution obtained using Equations \eqref{CauchyProp2} and \eqref{conditions}, so confirming the validity of the procedure.  

\subsection{Transmission spectra for the FEM model and the relaxed micromorphic continuum}

\begin{figure}[H]
	\begin{centering}
			\begin{picture}(315,155)
	\put(0,0){	\includegraphics[width=10.5cm]{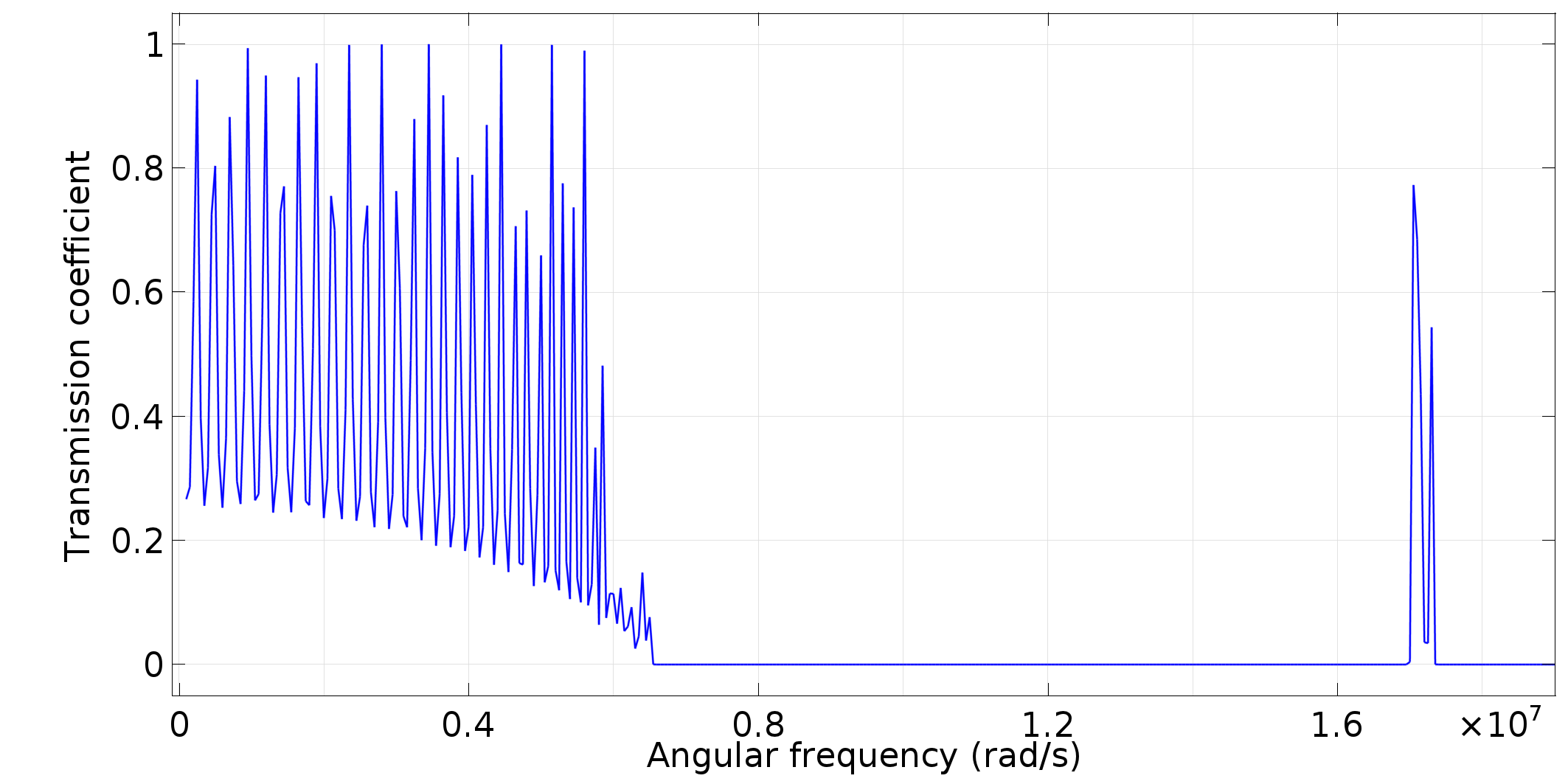}}
\end{picture}
		\par\end{centering}
\protect\caption{\label{fig:SpectrumLFEM} Transmission coefficient as
function of frequency for an incident longitudinal wave for the FEM model.}	
\end{figure}

\begin{figure}[H]

\begin{centering}
			\begin{picture}(315,155)
\put(9,0){		\includegraphics[width=10.1cm]{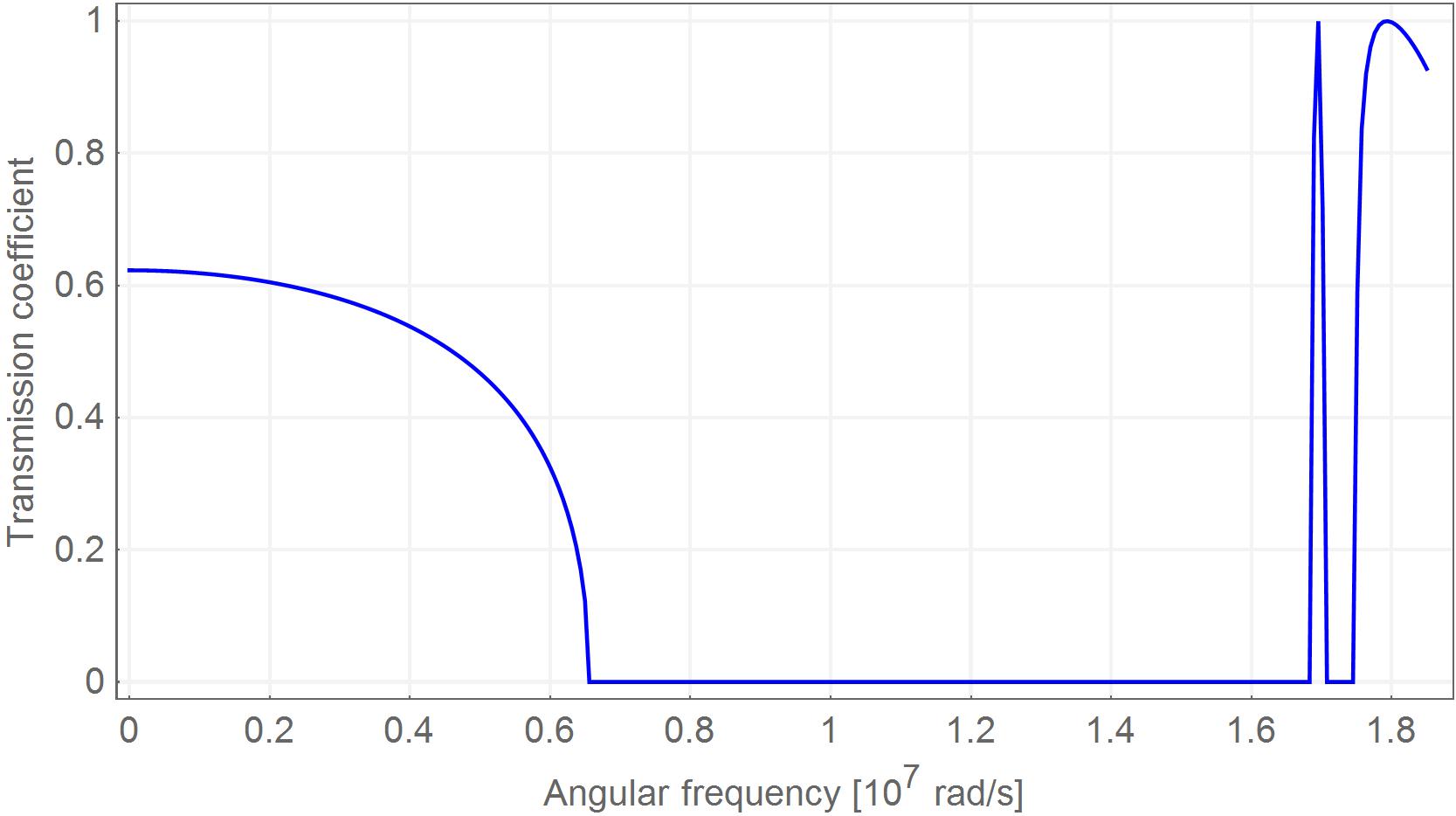}}
\end{picture}

		\par\end{centering}
	\protect\caption{\label{fig:SpectrumLREL} Transmission coefficient as
		function of frequency for an incident longitudinal wave for the relaxed micromorphic model.}	
\end{figure}

In this subsection, we compare the resulting transmission spectra for both the FEM model, as computed with the semi-analytical method proposed in section \ref{Spectra}, and the relaxed micromorphic one. In Figures \ref{fig:SpectrumLFEM} and \ref{fig:SpectrumLREL}, we show the spectra obtained considering a longitudinal traveling wave arriving at the interface for the FEM model and for the relaxed micromorphic continuum, respectively. 


\begin{figure}[H]
	\begin{centering}
		\includegraphics[width=10.5cm]{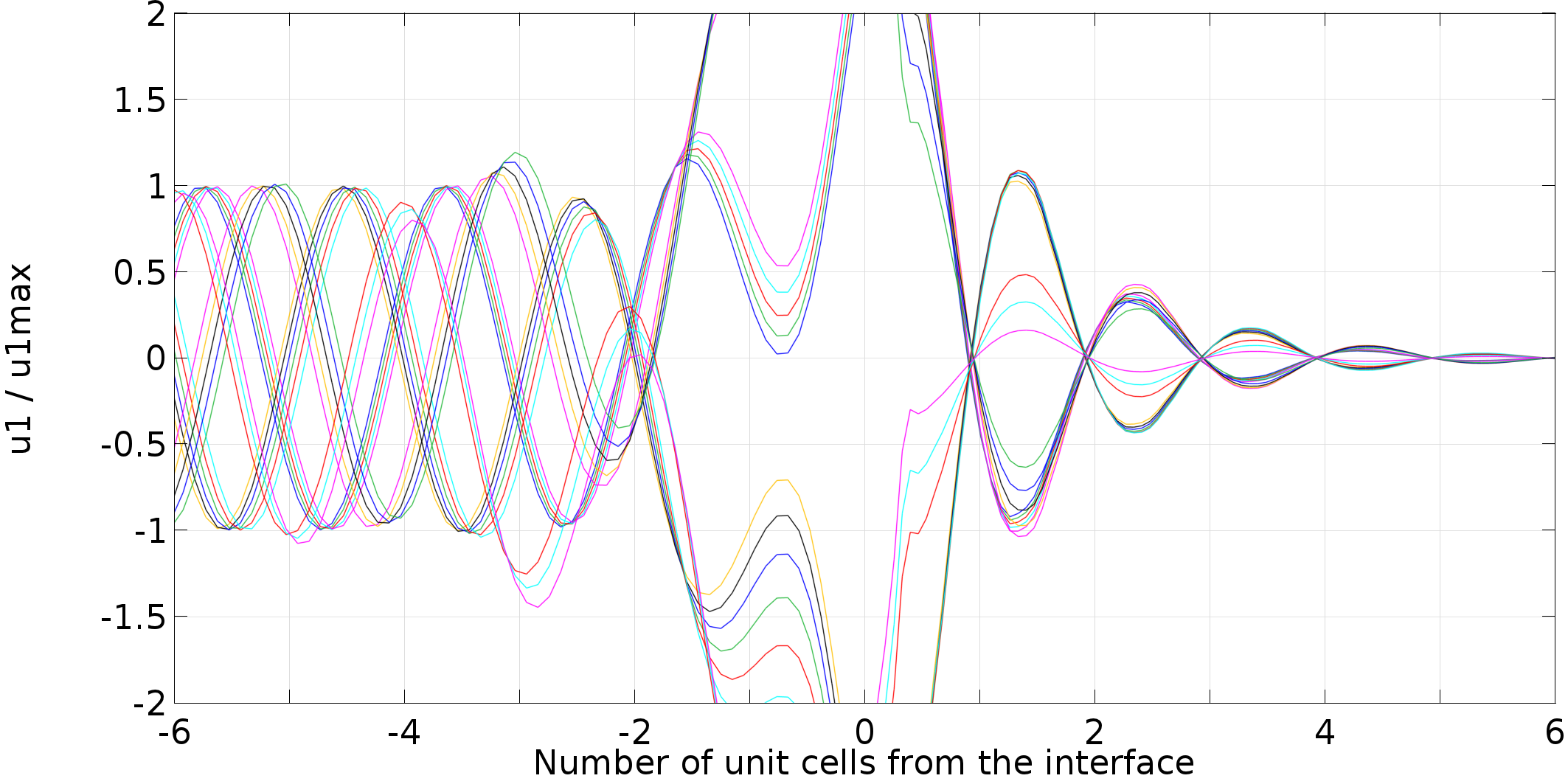}
				\par\end{centering}
	\protect\caption{\label{fig:ModeReflection1} Dimensionless displacement on the bottom edge for the vibrating modes in the range of angular frequencies $1.75\times 10^7$--$1.9\times 10^7$ rad/s (FEM simulations).}	
\end{figure}

\begin{figure}[H]
	\begin{centering}
		\includegraphics[width=10.5cm]{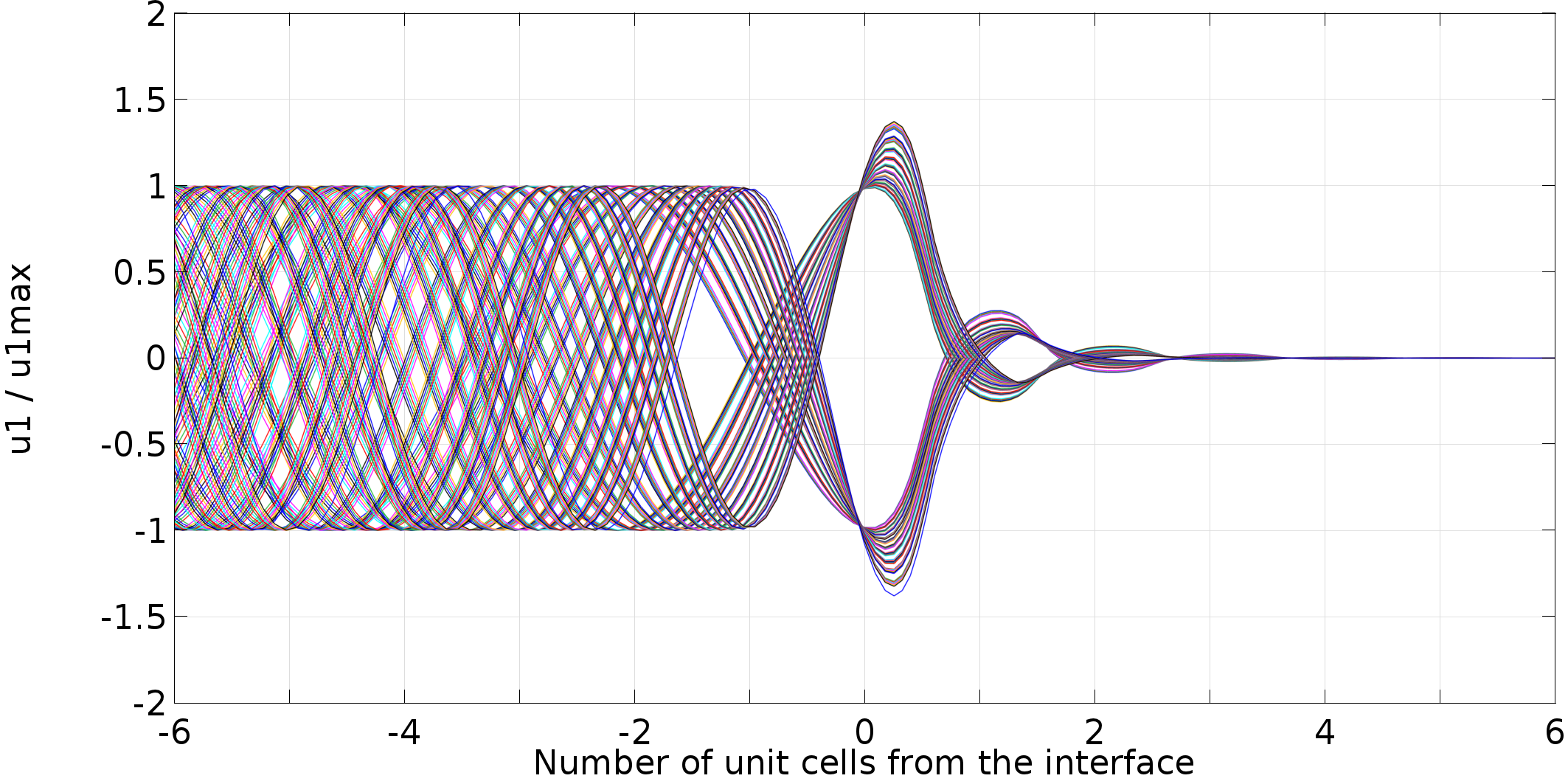}
		\par\end{centering}
	\protect\caption{\label{fig:ModeReflection2} Dimensionless displacement on the bottom edge for the vibrating modes in the range of angular frequencies  $0.80\times 10^7$--$1.5\times 10^7$ rad/s (FEM simulations).}	
\end{figure}

It is possible to see that an accurate average description of the effects is obtained. The transmission coefficient starts around 0.6 and becomes zero at around $0.7 \times  10^7$ rad/s in both the models and the first peak, even if higher in the relaxed micromorphic model, is comparable. Actually, the height of the first peak in Figure \ref{fig:SpectrumLFEM} depends on the frequency step of the calculation and can be more effectively described by choosing smaller frequency-steps close to the point of interest. In this respect, the relaxed model shows a better performance, since the analytical solution found for the relaxed model does not need to be discretized. For higher frequencies, the relaxed micromorphic model shows a second peak, while the FEM does not seem to allow any transmission. It is also useful to point out that, in the FEM model,  the propagation of the wave through the interface is somehow present but the amplitude of the wave decreases inside the microstructured metamaterial becoming zero after approximately 7 unit cells, see Figure \ref{fig:ModeReflection1}. On the other hand, for the range of frequencies of the central band gap the amplitude of the displacements becomes zero near the interface and the reflection can be entirely attributed to the presence of the interface (see Figure \ref{fig:ModeReflection2}). This difference seems to indicate that the transmission at the interface for the second optic mode happens but the wave continues to reflect inside the microstructured material.

As a matter of fact, for the frequencies considered in Figure \ref{fig:ModeReflection1}, the corresponding wavelengths of the incident wave start to become very small with respect to the size of the cell, so that Bragg scattering is sensible to take place. This fact is somehow captured by the FEM model, but cannot be captured by the relaxed model which is intrinsically a continuum model. In this case the frequency is too high in order to ensure the hypothesis of continuum model is still completely representative of reality.


\begin{figure}[H]
	\begin{centering}
					\begin{picture}(315,155)
		\put(0,0){
		\includegraphics[width=10.5cm]{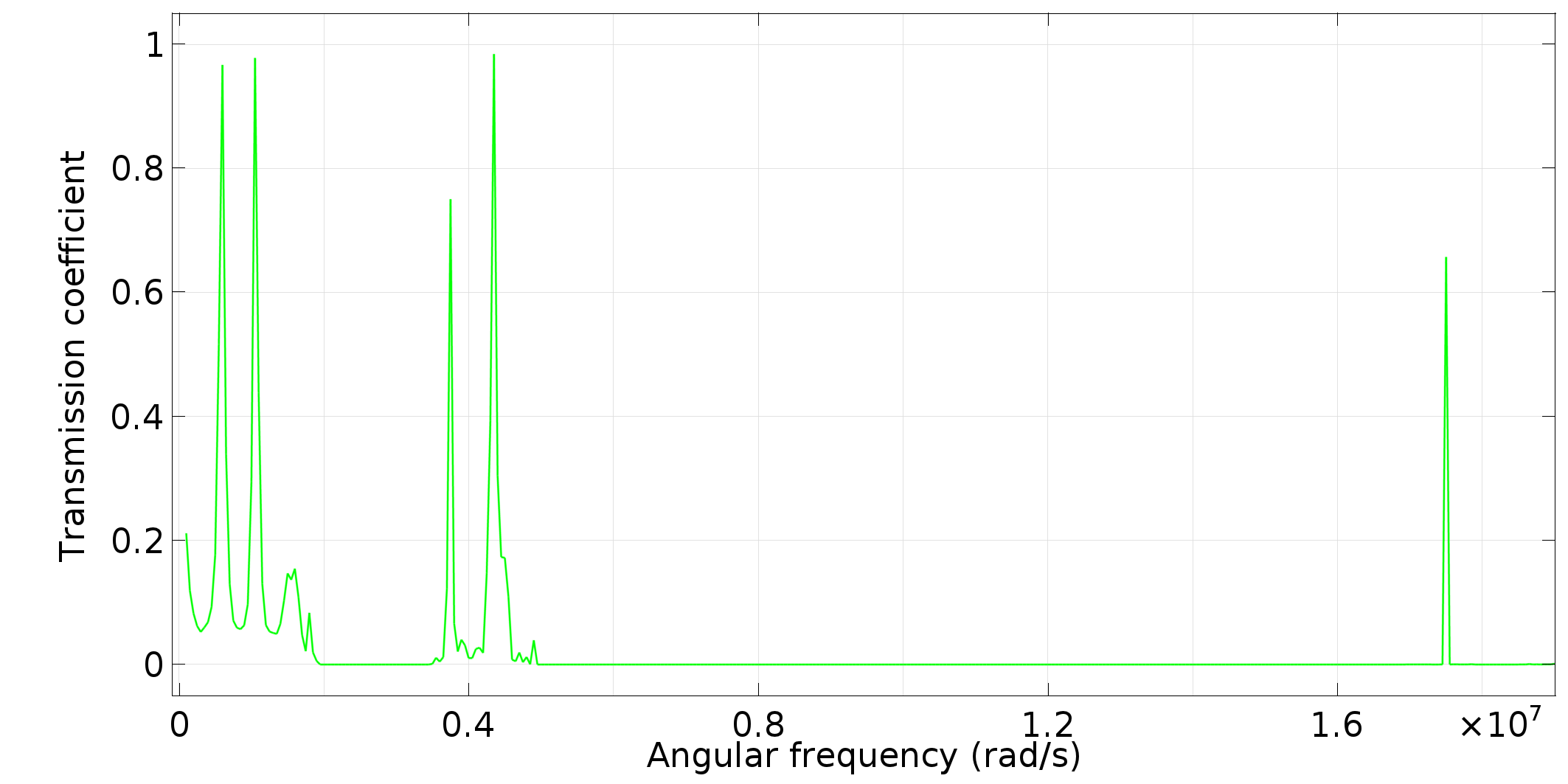}
	}
	\end{picture}
	
	\par\end{centering}
		\protect\caption{\label{fig:SpectrumTFEM}Transmission coefficient as function of frequency for an incident transverse wave for the FEM model.}	
\end{figure}

\begin{figure}[H]
	
	\begin{centering}
\begin{picture}(315,155)
\put(9,0){	 
			\includegraphics[width=10.1cm]{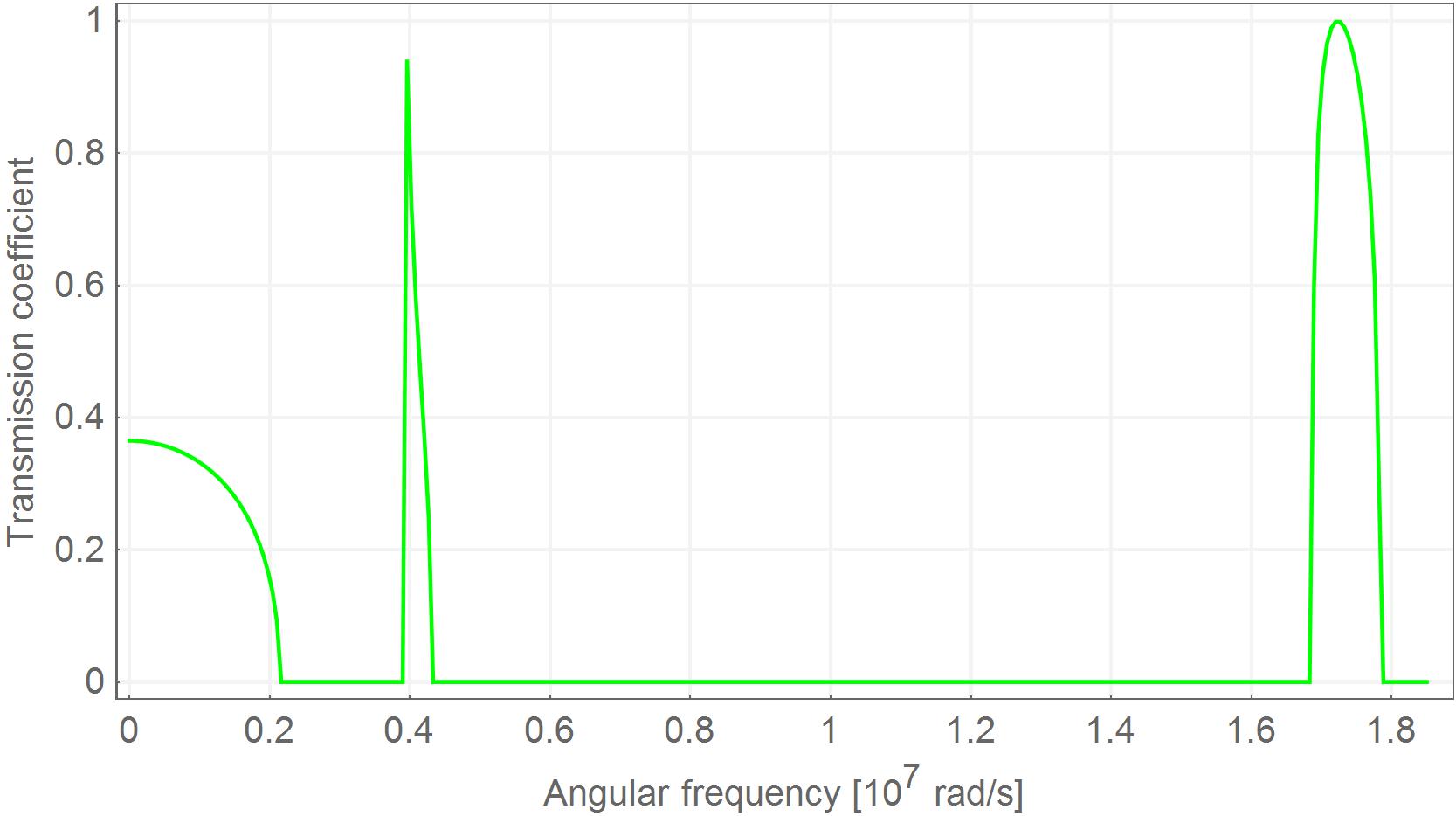}}
		\end{picture}
		\par\end{centering}
	\protect\caption{\label{fig:SpectrumTREL}Transmission coefficient as function of frequency for an incident transverse wave for the relaxed micromorphic model.}	
\end{figure}

The same analysis can be done for the transverse waves, as shown in Figures \ref{fig:SpectrumTFEM} and \ref{fig:SpectrumTREL}. In this case, the approximation given by the continuous model is even better because no extra peak can be found for higher frequencies. The two transmission peaks are fully described even if the value of the transmission coefficient is not exactly analogous. As before, the height of the peaks is better caught by the FEM model when adding more frequency points. Also, the width of the peaks is comparable with the only exception of the second one. 


\section{Conclusions}

In this paper, we provide a validation of the material parameters of the relaxed micromorphic model derived in \cite{madeo2016modeling} by means of the study of the transmission properties of a rather simplified meta-structure made up of 40 unit cells of a metamaterial made of periodic cross-like holes  \cite{madeo2016modeling}. In particular, the interface between a homogeneous solid and this meta-structure is considered and the reflection and transmission coefficients are derived as a function of the frequency.

The transmission spectra are computed both via a direct FEM simulation and via a direct implementation of the relaxed model with the material constants derived in \cite{madeo2016modeling}. The obtained results show an excellent agreement and the relaxed model revealed to be more than 10 times faster in terms of computational time with respect to the FEM implementation of the same problem (1-2 minutes for the relaxed model against 20 minutes for the FEM).

The present paper represents the first step towards the use of the relaxed micromorphic model for the characterization of the mechanical behavior of metamaterials and for their use in view of meta-structural design in the simplified framework of enriched continuum mechanics.

Future work will be focused on the generalization of the results presented here to the anisotropic framework in order to be able to characterize a wider class of metamaterials, so increasing the interest of using enriched continuum models for realistic meta-structural design.

\section{Acknowledgments}

Angela Madeo thanks INSA-Lyon for the funding of the BQR 2016 \textquotedbl{}Caract\'erisation
m\'ecanique inverse des m\'etamat\'eriaux: mod\'elisation, identification
exp\'erimentale des param\'etres et \'evolutions possibles\textquotedbl{},
as well as the CNRS-INSIS for the funding of the PEPS project.
Marco Miniaci acknowledges funding from the European Union's Horizon 2020 research and innovation programme under the Marie Sklodowska-Curie grant agreement no. 658483

{\footnotesize{}\let\stdsection\section \def\section *#1{\stdsection{#1}}}{\footnotesize \par}

{\footnotesize{}\bibliographystyle{plain}
\bibliography{library}
 \bibliographystyle{plain}}{\footnotesize \par}

{\footnotesize{}\let\section\stdsection}{\footnotesize \par}

\setcounter{section}{0}
\renewcommand\thesection{\Alph{section}}

\section
{Appendix - Derivation of the energy flux}

To prove that in relaxed micromorphic media the energy flux takes
the form \eqref{Flux}, we start noticing that, using eqs. \eqref{Energy}
and \eqref{Kinetic}, the time derivative of the total energy $E$
can be computed as 
\begin{align}
E_{,t}= & \:\rho\left\langle u_{,t},u_{,tt}\right\rangle +\eta\left\langle \p_{,t},\p_{,tt}\right\rangle +
\left\langle \overline{\eta}_{1} \dev\sym\nablau_{,tt},\dev\sym\nablau_{,t}\right\rangle
+\left\langle \overline{\eta}_{2} \skew\nablau_{,tt}, \skew\nablau_{,t}\right\rangle\nonumber\\ &
+\left\langle \overline{\eta}_{3} \tr\left(\nablau_{,tt}\right) \id ,\nablau_{,t}\right\rangle
+\left\langle 2\,\me\,\mathrm{sym}\left(\nablau-\p\right),\,\mathrm{sym}\left(\nablau_{,t}-\p_{,t}\right)\right\rangle \vspace{1.2mm}\nonumber\\
& +\left\langle \lle\mathrm{tr}\left(\nablau-\p\right)\mathds1,\,\nablau_{,t}-\p_{,t}\right\rangle +\left\langle 2\,\mu_{c}\,\mathrm{skew}\left(\nablau-\p\right),\,\mathrm{skew}\left(\nablau_{,t}-\p_{,t}\right)\right\rangle \vspace{1.2mm}\\
& +\left\langle 2\,\mu_{\rm micro}\:\mathrm{sym}\,\p,\mathrm{sym}\,\p_{,t}\right\rangle +\left\langle \lambda_{\rm micro}\,(\mathrm{tr}\,\p)\,\mathds1,\,\p_{,t}\right\rangle +\left\langle \mLc\,\mathrm{Curl}\,\p,\,\mathrm{Curl}\,\p_{,t}\right\rangle ,\nonumber
\end{align}
or equivalently, using definitions \eqref{quantities}
for $\mathcal{I}$, $\widetilde{\sigma}$, $s$ and $m$
\begin{align}
E_{,t}=\: & \rho\left\langle u_{,t},u_{,tt}\right\rangle +\eta\left\langle \p_{,t},\p_{,tt}\right\rangle+  
\left\langle \overline{\eta}_{1} \dev\sym\nablau_{,tt}+\overline{\eta}_{2} \skew\nablau_{,tt}+\overline{\eta}_{3} \tr\left(\nablau_{,tt}\right) \id ,\nablau_{,t}\right\rangle
\vspace{1.2mm}\nonumber\\
&+\left\langle 2\,\me\,\mathrm{sym}\left(\nablau-\p\right)+\lle\,\mathrm{tr}\left(\nablau-\p\right)\mathds1+2\,\mu_{c}\,\mathrm{skew}\left(\nablau-\p\right),\,\nablau_{,t}-\p_{,t}\right\rangle \vspace{1.2mm}\nonumber\\
& +\left\langle 2\,\mu_{\rm micro}\,\mathrm{sym}\,\p+\lambda_{\rm micro}\:(\mathrm{tr}\,\p)\mathds1,\,\p_{,t}\right\rangle +\left\langle \mLc\,\mathrm{Curl}\,\p,\,\mathrm{Curl}\,\p_{,t}\right\rangle\vspace{1.2mm}\\
=\: & \rho\left\langle u_{,t},u_{,tt}\right\rangle +\eta\left\langle \p_{,t},\p_{,tt}\right\rangle +\left\langle \widetilde{\sigma}+\mathcal{I},\nablau_{,t}\right\rangle -\left\langle \widetilde{\sigma\,}-s\,,\p_{,t}\right\rangle +\left\langle m\,,\,\mathrm{Curl}\,\p_{,t}\right\rangle .\nonumber
\end{align}
Using now the equations of motion \eqref{eq:Dyn} to replace
the quantities $\rho\,u_{,tt}$ and $\eta\,P_{,tt}$, recalling that
$\left\langle m\,,\,\mathrm{Curl}\,\p_{,t}\right\rangle =\mathrm{Div}\left(\,\left(m\,^{T}\cdot\p_{,t}\right):\epsilon\,\right)-\left\langle \mathrm{Curl}\,m\,,\,\p_{,t}\right\rangle $,
manipulating and simplifying, it can be recognized that 
\begin{align}\nonumber
E_{,t} & =\left\langle u_{,t},\mathrm{Div} \left[\widetilde{\sigma}+\mathcal{I}\right]\right\rangle +\left\langle \p_{,t},\widetilde{\sigma}-s\,-\mathrm{Curl}\,m\,\right\rangle +\mathrm{Div}\left(u_{,t}\cdot\left[\widetilde{\sigma}+\mathcal{I}\right]\right)-\left\langle u_{,t},\mathrm{Div}\left[\widetilde{\sigma}+\mathcal{I}\right]\right\rangle \vspace{1.2mm}\\
&\quad  -\left\langle \widetilde{\sigma}-s\,,\p_{,t}\right\rangle +\left\langle m\,,\,\mathrm{Curl}\,\p_{,t}\right\rangle\\\nonumber &=\mathrm{Div}\left(\left[\widetilde{\sigma}+\mathcal{I}\right]^{T}\cdot u_{,t}\right)-\left\langle \mathrm{Curl}\,m\,,\,\p_{,t}\right\rangle +\mathrm{Div}\left(\,\left(m\,^{T}\cdot\p_{,t}\right):\epsilon\,\right)+\left\langle \mathrm{Curl}\,m\,,\,\p_{,t}\right\rangle \vspace{1.2mm}\\
& =\mathrm{Div}\left(\,\left[\widetilde{\sigma}+\mathcal{I}\right]^{T}\cdot u_{,t}+\left(m\,^{T}\cdot\p_{,t}\right):\epsilon\,\right).\nonumber
\end{align}

\end{document}